\documentclass[]{aa}  
\usepackage{epsfig}
\usepackage{wasysym}
\usepackage{graphicx}
\usepackage{natbib}
\usepackage{booktabs}
\usepackage{lscape}
\usepackage{url}
\usepackage[breaklinks]{hyperref}

\usepackage{graphicx}
\usepackage{txfonts}
  \def \teff {$T_{\mathrm{eff}}$}
  \def \tc {$T_{\mathrm{c}}$}
  \def \vtur {$V_{\mathrm{tur}}$}
  \def \logg {$\log g$}
  \def \rmean {$R_{\mathrm{mean}}$}
  \def \zettwo {$\zeta^2$ Ret}
  \def \zetone {$\zeta^1$ Ret}

\begin{document}

   \title{Abundance trend with condensation temperature for stars with different Galactic birth places.\thanks{Based on 
   observations collected with the HARPS spectrograph
    at the 3.6-m telescope (program ID: 095.D-0717(A)), installed at the La Silla
    Observatory, ESO (Chile), with the UVES spectrograph at the 8-m Very
    Large Telescope (program ID: 095.D-0717(B)), installed at the Cerro Paranal Observatory, ESO (Chile).
    Also based on data obtained from the ESO Science Archive Facility under request numbers: vadibekyan180760, vadibekyan180762, 
    vadibekyan180764, vadibekyan180768, vadibekyan180769, vadibekyan180771, vadibekyan180773, vadibekyan180778, and vadibekyan180779.}
    \thanks{The Tables with stellar parameters and chemical abundances are available in electronic form at the CDS via anonymous ftp to cdsarc.u-strasbg.fr (130.79.128.5)
or via http://cdsweb.u-strasbg.fr/cgi-bin/qcat?J/A+A/}}

    \titlerunning{\tc \ trend and \rmean}

  \author{V. Adibekyan\inst{1}  
          \and E. Delgado-Mena\inst{1}
          \and P. Figueira\inst{1}
          \and S. G. Sousa\inst{1}
          \and N. C. Santos\inst{1,2}
          \and J.~I.~Gonz\'{a}lez Hern\'{a}ndez\inst{3,4}
          \and \\ I.~Minchev\inst{5}          
          \and J.~P.~Faria\inst{1,2}
          \and G. Israelian\inst{3,4}
          \and G. Harutyunyan\inst{5} 
          \and L. Su\'{a}rez-Andr\'{e}s\inst{3,4}
          \and A. A. Hakobyan\inst{6}
         }

  \institute{
          Instituto de Astrof\'isica e Ci\^encias do Espa\c{c}o, Universidade do Porto, CAUP, Rua das Estrelas, 4150-762 Porto, Portugal\\
          \email{Vardan.Adibekyan@astro.up.pt}
          \and
          Departamento de F\'isica e Astronomia, Faculdade de Ci\^encias, Universidade do Porto, Rua do Campo Alegre, 4169-007 Porto, Portugal
          \and 
          Instituto de Astrof\'{\i}sica de Canarias, 38200 La Laguna, Tenerife, Spain
          \and 
          Departamento de Astrof{\'\i}sica, Universidad de La Laguna, 38206 La Laguna, Tenerife, Spain        
          \and 
          Leibniz Institute for Astrophysics Potsdam (AIP), An der Sternwarte 16, 14482 Potsdam, Germany
          \and
          Byurakan Astrophysical Observatory, 0213 Byurakan, Aragatsotn province, Armenia
}

   \date{Received date / Accepted date}

% \abstract{}{}{}{}{} 
% 5 {} token are mandatory
 
  \abstract
   {During the past decade, several studies reported a correlation between chemical abundances of stars and condensation temperature (also known as \tc \ trend). However,
  the real astrophysical nature of this correlation is still debated.}
   {The main goal of this work is to explore the possible dependence of the \tc \ trend on stellar Galactocentric distances, \rmean.}
   {We used high-quality spectra of about 40 stars observed with the HARPS and UVES spectrographs to derive  precise stellar parameters, chemical abundances,
   and stellar ages. A differential line-by-line analysis was applied to achieve the highest possible precision in the chemical abundances.}
   {We confirm previous results that [X/Fe] abundance ratios depend on stellar age and that for a given age, some elements  also show a dependence on 
   \rmean. When using the whole sample of stars, we observe a weak hint that the \tc \ trend depends on \rmean. The observed dependence
   is very complex and disappears when only stars with similar ages are considered.}
   {To conclude on the possible dependence of the \tc \ trend on the formation place of stars, a larger sample of stars with very similar 
   atmospheric parameters and stellar ages observed at different Galactocentric distances is needed.}


   \maketitle
%
%________________________________________________________________________________________________________

\section{Introduction}                                  \label{sec:intro}

Stars and planets form from the same material, and as such, some of their properties are expected to be inter-connected. The very first correlation observed
in the field of exoplanet research was the dependence of giant-planet occurrence and stellar metallicity \citep[e.g.][]{Gonzalez-97, Santos-01, Santos-04, Fischer-05}.
This correlation eventually played one of the most important roles on our understanding and constraining of planet formation 
theories \citep[e.g.][]{Mordasini-09}. Following works showed that the occurrence of different types of planets also may depend on chemical abundances
of the hosting stars \citep[e.g.][]{Haywood-08, Adibekyan-12a, Adibekyan-12b}.

The importance of stellar metallicity and chemical properties of stars is not limited to the formation of planets alone. 
The chemical abundances and some specific abundance ratios of stars with planets provide enormous amounts of information about 
the planet evolution process \citep[e.g.][]{Bond-10, Delgado-10}, architecture \citep[e.g.][]{Dawson-13, Beauge-13, Adibekyan-13},
composition of planets \citep[e.g.][]{Thiabaud-14, Dorn-15, Santos-15}, and even about the habitability of planets \citep[e.g.][]{Adibekyan-16}.

During the past decades, astronomers have also been searching for chemical signatures of planet formation on the planet-host stars. 
Many studies, starting from \citet{Gonzalez-97} and \citet{Smith-01}, explored a possible trend between the abundances of chemical elements and 
the condensation temperature (\tc) of the elements to understand the relative fraction of volatile (low-\tc)  and refractory (high-\tc) elements 
in planet-host and single stars. This trend is called \tc \ trend, and the slope of the correlation (slope of the linear fit) of [X/Fe] vs. condensation 
temperature is named \tc \ slope.

\citet[][]{Melendez-09} was the first to report that the Sun shows a deficit in refractory elements with respect to other solar twins. They suggested that this is 
due to these elements being trapped in the terrestrial planets in our solar system. Although the authors discussed the effect of the Galactic evolution 
\citep[taking into account the possibility that the Sun might have migrated from an inner Galactic orbit,  e.g.][]{Wielen-96}, they finally did
not consider it a plausible explanation. 
The same conclusion (presence of planet formation signatures in the atmospheres of stars) was  also reached by 
\citet{Ramirez-09}, who analyzed 64 solar twins and analogues. However, the results by \citet{GH-10} and \citet{GH-13} strongly contested the connection between
the presence of planets and the abundance peculiarities of the stars. This very exciting and important (possible) connection  between chemical 
peculiarities and formation of planets has also been examined in other works \citep[e.g.][]{Takeda-01, Ecuvillon-06, Sozzetti-06, 
Schuler-11, GH-13, Maldonado-15, Nissen-15, Biazzo-15, Saffe-15, Saffe-16}, but contradictory conclusions were reached.

The fact that the observed \tc \  trend may be a relic of the planet formation and evolution process is by far not its only possible explanation.
Recently, \citet{Adibekyan-14} used a sample of 148 solar-type 
stars from \citet{GH-10} and \citet{GH-13} to explore the possible factors responsible for the \tc \ trend.
The authors found that the slope of this trend  correlates with the stellar age in a significant way: more evolved (old) stars have a lower refractory-to-volatile
ratio. Since we do not expect significant changes of chemical abundances  with age for FGK dwarf stars in the main sequence, \citet{Adibekyan-14} concluded that the 
observed correlation 
probably reflects the chemical evolution in the Galaxy. \citet{Ramirez-14} also observed a correlation between the \tc \ slope and stellar age
for metal-rich solar analogues, but the sign of the correlation seemed to be opposite to what was obtained in \citet{Adibekyan-14}\footnote{We note that the authors studied the abundances of the stars
relative to the pristine sample (stars with the highest refractory-to-volatile ratio), while our abundances are relative to the Sun.}. %
They found that most refractory-element-depleted stars are younger than those with the highest refractory element abundances. In
addition to these 
explanations, it was suggested that the \tc \ trend strongly correlates with stellar radius and mass \citep{Maldonado-15, Maldonado-16}, 
it may also depend on stellar environment \citep{Onehag-14} and internal processes, such as gas-dust segregation in the protostellar disk \citep{Gaidos-15}.

\citet{Adibekyan-14} found tentative evidence that the \tc \ slopes also correlate  with the 
mean Galactocentric distance of the stars (\rmean), indicating that stars that originated in the inner Galaxy have fewer refractory elements than volatiles.  Their sample was composed of stars with different ages and was also lacking stars with
low \rmean \ values (most of the stars were clustered at \rmean \ $\approx$ 7-8 kpc). It was 
difficult to  firmly conclude about the direct role of \rmean \ in the observed correlation between \tc \ slope and \rmean\
based on this sample alone. 

In this context, the study of binary stars is important because the above mentioned mechanisms and processes cannot easily explain trends observed between companions 
of these systems. Several authors studied the \tc \ trend in binary stars with and without planetary
companions \citep[e.g.][]{Liu-14, Saffe-15, Mack-16} or in binary stars where both components host planets \citep[e.g.][]{Biazzo-15, Teske-15, Ramirez-15, Teske-16}.
The results and conclusions of these studies point in different directions and show as a whole that there are no systematic differences in the chemical abundances of 
stars with and without planets in the binary systems. Moreover, there are discrepancies in the results even for the same individual systems such as 16 Cyg AB
\citep[e.g. ][]{Laws-01, Takeda-05, Schuler-11a, TucciMaia-14}. 

Very recently, \citet[][]{Saffe-16} reported a positive \tc \ trend in the binary system \zetone \ -- \zettwo, \ where one of the stars (\zettwo) hosts a debris disk. 
The deficit of the refractory elements relative to volatiles in \zettwo \ was explained by the authors as being caused by the depletion of 
about $\sim$3M$_{\oplus}$  rocky material. Following their work, \citet{Adibekyan-16a} confirmed
the trend observed between the stars based on a very high quality
data set. However, they also found that the \tc \ trends seem to depend  on the individual spectrum used (even if always of very high quality). 
In particular, they observed significant differences in the abundances of the same star derived from different high-quality spectra.

In this work, we explore the possible dependence of the \tc \ trend on Galactocentric distances for a fixed formation time, that is, for stars with very similar ages.
We organized this paper as follows. In Sect.\,\ref{sec:data} we  present the data, in Sects.\,\ref{sec:parameters} and \ref{sec:ages}
we present the methods used to derive the stellar parameters, chemical abundances, and stellar ages.  
Section \,\ref{sec:tc_slope} explains how we calculate and evaluate the significance of the \tc \ trends. 
After presenting the main results in Sect.\,\ref{sec:tc_rmean}, we  conclude in Sect.\,\ref{sec:conclusion}.

\begin{table*}[t!]

\setlength{\tabcolsep}{3pt}
\caption{\label{tab:parameters}  Spectroscopic observations and stellar parameters, ages, and \tc \ slopes of the sample stars.}
\begin{center}
\begin{tabular}{lcccccccc}
\hline\hline
Star & \teff & $\log g$  & [Fe/H]  & \vtur \  & Age  & \tc \ slope  & S/N & Instrument \\
 & (K) & dex) & (dex) & (km s$^{-1}$) &  (Gyr) & (10$^{-4}$ dex K$^{-1}$) & &  \\
\hline
HD8828 & 5380$\pm$19 & 4.39$\pm$0.03 & -0.16$\pm$0.01 & 0.72$\pm$0.04 & 7.1$\pm$3.6 & 0.11$\pm$0.33 & 980 & HARPS \tabularnewline
HD11271 & 6120$\pm$21 & 4.37$\pm$0.03 & 0.23$\pm$0.02 & 1.24$\pm$0.02 & 2.7$\pm$0.6 & 0.06$\pm$0.27 & 380 & HARPS \tabularnewline
HD23079 & 5980$\pm$13 & 4.46$\pm$0.03 & -0.12$\pm$0.01 & 1.10$\pm$0.02 & 5.4$\pm$0.5 & 0.65$\pm$0.30 & 1080 & HARPS \tabularnewline
HD44594 & 5860$\pm$17 & 4.40$\pm$0.02 & 0.17$\pm$0.01 & 1.02$\pm$0.02 & 2.2$\pm$0.7 & 0.50$\pm$0.17 & 1170 & HARPS \tabularnewline
HD45067 & 6100$\pm$44 & 4.16$\pm$0.03 & 0.02$\pm$0.03 & 1.35$\pm$0.06 & 3.9$\pm$0.6 & 0.82$\pm$0.37 & 390 & UVES \tabularnewline
HD59967 & 5820$\pm$18 & 4.48$\pm$0.03 & -0.05$\pm$0.01 & 1.17$\pm$0.03 & 1.0$\pm$0.7 & 1.29$\pm$0.60 & 420 & HARPS \tabularnewline
HD69830 & 5390$\pm$23 & 4.39$\pm$0.04 & -0.05$\pm$0.02 & 0.73$\pm$0.04 & 9.3$\pm$1.6 & 0.57$\pm$0.21 & 1560 & HARPS \tabularnewline
HD77462 & 6321$\pm$35 & 4.38$\pm$0.03 & -0.19$\pm$0.02 & 1.38$\pm$0.05 & 3.9$\pm$0.5 & 0.36$\pm$0.34 & 510 & HARPS \tabularnewline
HD86997 & 6300$\pm$33 & 4.29$\pm$0.03 & -0.03$\pm$0.02 & 1.46$\pm$0.05 & 3.2$\pm$0.4 & 0.12$\pm$0.29 & 440 & HARPS \tabularnewline
HD90774 & 6150$\pm$28 & 4.25$\pm$0.03 & 0.02$\pm$0.02 & 1.37$\pm$0.03 & 3.5$\pm$0.5 & 0.50$\pm$0.30 & 400 & HARPS \tabularnewline
HD101198 & 6230$\pm$21 & 4.27$\pm$0.03 & -0.15$\pm$0.01 & 1.52$\pm$0.03 & 4.3$\pm$0.5 & -0.05$\pm$0.14 & 2350 & HARPS \tabularnewline
HD105665 & 6110$\pm$20 & 4.28$\pm$0.03 & -0.25$\pm$0.02 & 1.36$\pm$0.03 & 6.7$\pm$0.3 & -0.51$\pm$0.42 & 390 & HARPS \tabularnewline
HD106200 & 6490$\pm$113 & 4.72$\pm$0.06 & -0.26$\pm$0.07 & 2.45$\pm$0.29 & 3.1$\pm$1.0 & 0.34$\pm$0.61 & 180 & UVES \tabularnewline
HD106869 & 6130$\pm$17 & 4.34$\pm$0.04 & 0.14$\pm$0.01 & 1.27$\pm$0.02 & 3.1$\pm$0.2 & 0.24$\pm$0.15 & 430 & HARPS \tabularnewline
HD109591 & 5720$\pm$16 & 4.29$\pm$0.02 & -0.07$\pm$0.01 & 0.97$\pm$0.03 & 11.0$\pm$0.5 & -0.74$\pm$0.55 & 410 & HARPS \tabularnewline
HD115031 & 5880$\pm$15 & 4.26$\pm$0.02 & -0.01$\pm$0.01 & 1.10$\pm$0.02 & 5.3$\pm$2.1 & -0.56$\pm$0.14 & 780 & HARPS \tabularnewline
HD116941 & 6300$\pm$37 & 4.36$\pm$0.03 & -0.06$\pm$0.03 & 1.42$\pm$0.05 & 3.1$\pm$0.5 & -0.05$\pm$0.45 & 390 & HARPS \tabularnewline
HD117190 & 6376$\pm$35 & 4.65$\pm$0.04 & 0.03$\pm$0.02 & 1.50$\pm$0.05 & 1.1$\pm$0.8 & 2.07$\pm$0.49 & 350 & HARPS \tabularnewline
HD117618 & 5970$\pm$18 & 4.37$\pm$0.02 & 0.03$\pm$0.01 & 1.09$\pm$0.02 & 4.9$\pm$0.5 & -0.07$\pm$0.18 & 630 & HARPS \tabularnewline
HD119758 & 5760$\pm$52 & 4.42$\pm$0.06 & 0.04$\pm$0.04 & 1.13$\pm$0.08 & 2.8$\pm$2.5 & 0.90$\pm$0.37 & 340 & UVES \tabularnewline
HD122194 & 5840$\pm$18 & 4.34$\pm$0.02 & 0.09$\pm$0.01 & 1.07$\pm$0.02 & 2.9$\pm$2.4 & -0.13$\pm$0.22 & 560 & HARPS \tabularnewline
HD122603 & 6080$\pm$23 & 4.09$\pm$0.03 & 0.09$\pm$0.02 & 1.36$\pm$0.02 & 3.7$\pm$0.5 & 0.41$\pm$0.18 & 440 & HARPS \tabularnewline
HD128760 & 6160$\pm$20 & 4.45$\pm$0.03 & 0.17$\pm$0.01 & 1.24$\pm$0.02 & 2.5$\pm$0.7 & -0.02$\pm$0.18 & 350 & HARPS \tabularnewline
HD129290 & 5970$\pm$75 & 4.30$\pm$0.19 & 0.03$\pm$0.06 & 1.08$\pm$0.10 & 4.8$\pm$2.2 & 0.72$\pm$0.33 & 140 & UVES \tabularnewline
HD130265 & 5640$\pm$52 & 4.34$\pm$0.06 & -0.21$\pm$0.04 & 0.76$\pm$0.09 & 7.3$\pm$3.8 & 0.68$\pm$0.42 & 270 & UVES \tabularnewline
HD134330 & 5620$\pm$23 & 4.46$\pm$0.03 & 0.10$\pm$0.02 & 0.95$\pm$0.04 & 1.5$\pm$1.3 & 0.97$\pm$0.23 & 420 & HARPS \tabularnewline
HD139590 & 6190$\pm$20 & 4.42$\pm$0.03 & 0.15$\pm$0.01 & 1.27$\pm$0.03 & 2.6$\pm$0.3 & -0.12$\pm$0.35 & 400 & HARPS \tabularnewline
HD142921 & 5850$\pm$14 & 4.30$\pm$0.02 & -0.08$\pm$0.01 & 1.05$\pm$0.02 & 9.0$\pm$0.3 & -1.88$\pm$0.56 & 380 & HARPS \tabularnewline
HD146546 & 6211$\pm$24 & 4.37$\pm$0.03 & -0.02$\pm$0.02 & 1.35$\pm$0.03 & 3.7$\pm$0.4 & 0.01$\pm$0.26 & 360 & HARPS \tabularnewline
HD147644 & 6040$\pm$18 & 4.38$\pm$0.02 & -0.10$\pm$0.01 & 1.19$\pm$0.02 & 2.1$\pm$1.5 & 0.37$\pm$0.19 & 360 & HARPS \tabularnewline
HD148998 & 6030$\pm$17 & 4.21$\pm$0.02 & 0.06$\pm$0.01 & 1.26$\pm$0.02 & 4.4$\pm$0.9 & 0.70$\pm$0.16 & 320 & HARPS \tabularnewline
HD157060 & 6270$\pm$25 & 4.39$\pm$0.04 & 0.09$\pm$0.02 & 1.38$\pm$0.03 & 2.7$\pm$0.2 & 0.70$\pm$0.74 & 470 & HARPS \tabularnewline
HD158469 & 6200$\pm$19 & 4.40$\pm$0.04 & 0.06$\pm$0.01 & 1.35$\pm$0.02 & 3.2$\pm$0.3 & 0.12$\pm$0.18 & 640 & HARPS \tabularnewline
HD168432 & 6180$\pm$26 & 4.36$\pm$0.03 & -0.19$\pm$0.02 & 1.32$\pm$0.03 & 5.5$\pm$0.4 & 1.00$\pm$0.31 & 410 & HARPS \tabularnewline
HD169822 & 5586$\pm$51 & 4.48$\pm$0.07 & -0.13$\pm$0.04 & 0.86$\pm$0.09 & 2.1$\pm$2.1 & -0.52$\pm$0.36 & 460 & HARPS \tabularnewline
HD175128 & 6110$\pm$22 & 4.31$\pm$0.04 & 0.10$\pm$0.02 & 1.29$\pm$0.03 & 3.4$\pm$1.0 & -0.06$\pm$0.15 & 330 & HARPS \tabularnewline
HD188345 & 6050$\pm$19 & 4.23$\pm$0.03 & 0.19$\pm$0.01 & 1.29$\pm$0.02 & 3.5$\pm$0.2 & 0.08$\pm$0.44 & 330 & HARPS \tabularnewline
HD198227 & 6190$\pm$22 & 4.30$\pm$0.03 & 0.08$\pm$0.02 & 1.33$\pm$0.03 & 3.2$\pm$0.4 & 0.39$\pm$0.31 & 270 & HARPS \tabularnewline
HD219272 & 6240$\pm$39 & 4.45$\pm$0.08 & 0.06$\pm$0.03 & 1.19$\pm$0.05 & 1.7$\pm$1.2 & 0.68$\pm$0.52 & 250 & UVES \tabularnewline
\hline
\end{tabular}
\end{center}
\end{table*}

\section{Data and observations}                                                 \label{sec:data}

To understand if the abundance trend observed with the condensation temperature is a function of Galactocentric distances for a fixed age of stars, 
we selected about 40 stars with ages similar to that of the Sun
but with different mean
Galactocentric distances from the Geneva-Copenhagen Survey sample \citep[GCS,][]{Nordstrom-04}: with the smallest (\rmean \ $\sim$6.5 kpc), largest (\rmean \ $\sim$9 kpc), and solar (\rmean \ $\sim$8 kpc) Galactocentric \rmean \ values\footnote{\cite{Casagrande-11} provides revised 
stellar parameters, ages, Galactic orbital parameters and the space velocity components for the GCS stars.}. %
The stars were selected to have \teff= T$_{\odot} \pm $500K, $\log g $ $>$ 4.0 dex, [Fe/H] = 0.0$\pm$0.2 dex, and an age of  4.6 $\pm$1Gyr
\footnote{We used BASTI expectation ages as suggested by \cite{Casagrande-11}.%
}. The range of \teff \ and $\log g $ were chosen (not very different from the solar values) to guarantee a high-precision differential chemical analysis.
However, we should note that because of the relatively wide range of stellar parameters, the achieved precision in chemical abundances is somewhat lower than
the precision obtained for solar twins and solar analogues \citep[e.g.][]{Melendez-09, Ramirez-09, GH-10, Nissen-15}.

High-resolution and high signal-to-noise (S/N) spectra for these stars were obtained by performing new observations with HARPS (22 stars) and UVES (six stars) 
ESO spectrographs, and by extracting spectra for 14 stars from the ESO archive\footnote{http://archive.eso.org/wdb/wdb/adp/phase3\_spectral/form}. 
One star was excluded from our analysis because of very low S/N (HD216054), HD213791 was excluded because of its very fast rotation, and HD184588 was excluded
because its  spectra are contaminated by the binary component. Our final sample thus contains 39 stars, three of which are known to harbour 
planets\footnote{http://exoplanets.eu/}. The three planet-host stars do not show any significant difference in stellar properties when compared to
the rest of the sample stars, thus we did not consider them as a separate group.

For the Sun we used a combined HARPS reflected spectrum from Vesta (extracted from the same public archive, S/N $\sim$1300).

\begin{table*}[t!]

\setlength{\tabcolsep}{3pt}
\caption{\label{tab:slopes} Slopes of the [X/H] versus metallicity, stellar age, and \rmean. A frequentist approach is chosen to derive 
the slopes and their uncertainties.}
\begin{center}
\begin{tabular}{lcccccc}
\hline\hline
Element &\multicolumn{2}{c}{[X/Fe] vs. [Fe/H]} & \multicolumn{2}{c}{[X/Fe] vs. Age} & \multicolumn{2}{c}{[X/Fe] vs. \rmean}\\ \cmidrule(l){2-3}\cmidrule(l){4-5}\cmidrule(l){6-7}
 & Slope$\pm \sigma$ & P(F-stat) & Slope$\pm \sigma$ & P(F-stat) & Slope$\pm \sigma$ & P(F-stat) \\
\hline
{[CI/Fe]} & -0.227$\pm$0.050 & 0.0001 & 0.019$\pm$0.006 & 0.0026 & 0.024$\pm$0.005 & 0.0001 \tabularnewline
{[OI/Fe]} & -0.312$\pm$0.079 & 0.0004 & 0.032$\pm$0.009 & 0.0016 & 0.027$\pm$0.011 & 0.0332 \tabularnewline
{[NaI/Fe]} & 0.036$\pm$0.044 & 0.422 & -0.003$\pm$0.005 & 0.5990 & -0.001$\pm$0.004 & 0.9130 \tabularnewline
{[MgI/Fe]} & -0.191$\pm$0.069 & 0.0085 & 0.018$\pm$0.004 & 0.0004 & 0.011$\pm$0.006 & 0.1020 \tabularnewline
{[AlI/Fe]} & -0.057$\pm$0.078 & 0.472 & -0.012$\pm$0.006 & 0.0413 & -0.022$\pm$0.006 & 0.0037 \tabularnewline
{[SiI/Fe]} & -0.086$\pm$0.039 & 0.0324 & 0.008$\pm$0.002 & 0.0018 & 0.002$\pm$0.003 & 0.5560 \tabularnewline
{[SI/Fe]} & -0.118$\pm$0.055 & 0.0376 & 0.006$\pm$0.006 & 0.3910 & 0.009$\pm$0.007 & 0.2550 \tabularnewline
{[CaI/Fe]} & -0.175$\pm$0.024 & 1.6$\times 10^{-8}$ & 0.006$\pm$0.004 & 0.1180 & 0.014$\pm$0.003 & 0.0001 \tabularnewline
{[ScII/Fe]} & 0.076$\pm$0.075 & 0.316 & 0.000$\pm$0.009 & 0.9950 & -0.026$\pm$0.008 & 0.0032 \tabularnewline
{[$<$Ti$>$/Fe]} & -0.183$\pm$0.039 & 4.0$\times 10^{-5}$ & 0.013$\pm$0.003 & 0.0008 & 0.009$\pm$0.004 & 0.0534 \tabularnewline
{[VI/Fe]} & -0.061$\pm$0.060 & 0.312 & -0.006$\pm$0.004 & 0.1670 & -0.005$\pm$0.006 & 0.4170 \tabularnewline
{[CrI/Fe]} & -0.042$\pm$0.019 & 0.0372 & -0.005$\pm$0.002 & 0.0679 & 0.000$\pm$0.003 & 0.9660 \tabularnewline
{[CoI/Fe]} & -0.038$\pm$0.062 & 0.539 & 0.011$\pm$0.005 & 0.0385 & -0.014$\pm$0.005 & 0.0169 \tabularnewline
{[NiI/Fe]} & 0.102$\pm$0.028 & 0.0008 & 0.001$\pm$0.004 & 0.8160 & -0.016$\pm$0.003 & 0.0002 \tabularnewline
{[CuI/Fe]} & -0.018$\pm$0.086 & 0.833 & -0.006$\pm$0.010 & 0.5710 & -0.024$\pm$0.009 & 0.0184 \tabularnewline
{[ZnI/Fe]} & -0.075$\pm$0.088 & 0.401 & 0.008$\pm$0.005 & 0.1530 & -0.013$\pm$0.004 & 0.0032 \tabularnewline
{[SrI/Fe]} & 0.013$\pm$0.096 & 0.894 & -0.013$\pm$0.007 & 0.0953 & 0.002$\pm$0.011 & 0.8330 \tabularnewline
{[YII/Fe]} & 0.192$\pm$0.086 & 0.0312 & -0.016$\pm$0.005 & 0.0061 & 0.008$\pm$0.006 & 0.2400 \tabularnewline
{[ZrII/Fe]} & -0.078$\pm$0.089 & 0.382 & 0.004$\pm$0.006 & 0.4970 & 0.026$\pm$0.006 & 0.0009 \tabularnewline
{[BaII/Fe]} & -0.195$\pm$0.101 & 0.0617 & -0.015$\pm$0.011 & 0.2070 & 0.018$\pm$0.008 & 0.0334 \tabularnewline
{[CeII/Fe]} & -0.270$\pm$0.084 & 0.003 & -0.019$\pm$0.009 & 0.0382 & 0.001$\pm$0.010 & 0.9270 \tabularnewline
{[NdII/Fe]} & -0.275$\pm$0.064 & 0.0001 & 0.020$\pm$0.011 & 0.0735 & 0.032$\pm$0.010 & 0.0044 \tabularnewline
\hline
\end{tabular}
\end{center}
\end{table*}

\section{Stellar parameters and chemical abundances}            \label{sec:parameters}

The stellar parameters (\teff, [Fe/H], $\log g$, and \vtur) and chemical abundances of the stars relative to the Sun were derived by the methods described in 
\citet{Adibekyan-16a}. In brief,  first we automatically measured the equivalent widths (EWs) of iron lines ($\sim$250 \ion{Fe}{i} and $\sim$40 \ion{Fe}{ii} lines) 
using the ARES v2 code\footnote{The last version of ARES code (ARES v2) can be downloaded at http://www.astro.up.pt/$\sim$sousasag/ares} \citep{Sousa-08, Sousa-15}.
Then the spectroscopic parameters were derived by imposing excitation and ionization balance assuming LTE. We used the grid of ATLAS9 plane-parallel model
of atmospheres \citep{Kurucz-93} and the 2014 version of the MOOG\footnote{The source code of MOOG can be downloaded at
http://www.as.utexas.edu/$\sim$chris/moog.html}  radiative transfer code \citep{Sneden-73}. 
The uncertainties of the parameters were derived as in our previous works, 
and they are well described in \citet[][]{Neuforge-97}. Stellar parameters of the stars are presented in Table\,\ref{tab:parameters}.
For more details on the derivation of stellar parameters we refer to \citet[][]{Sousa-14}. 

\begin{figure*}
\begin{center}
\begin{tabular}{c}
\includegraphics[angle=0,width=0.95\linewidth]{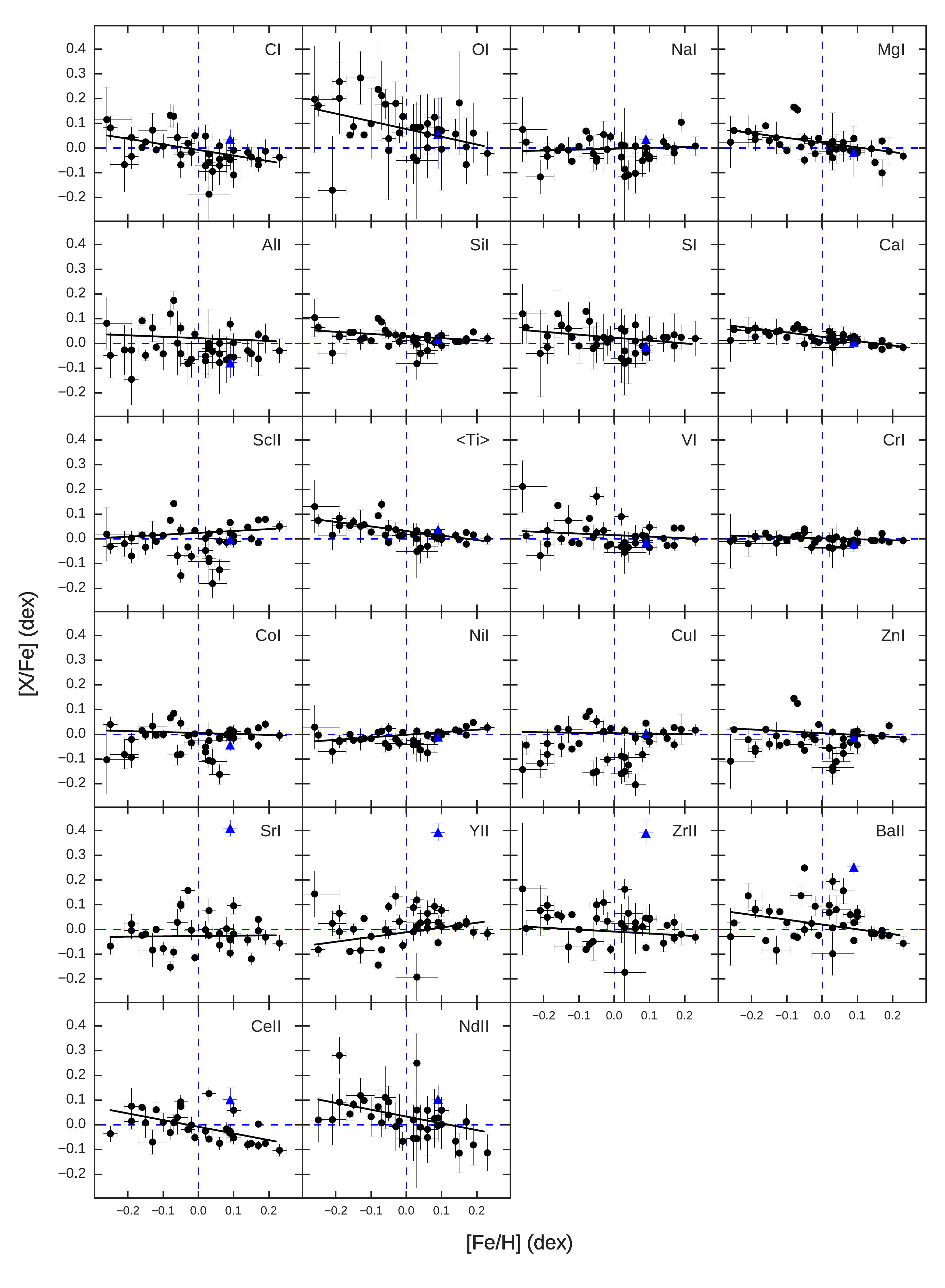}
\end{tabular}
\end{center}
\vspace{-0.9cm}
\caption{Abundance ratio [X/Fe] against [Fe/H] for the sample stars. The chemically peculiar
Ba star is shown by the blue triangle. The black line represents the WLS fit of the data (excluding the Ba star).}
\label{plot_elfe_feh}
\end{figure*}

\begin{figure*}
\begin{center}
\begin{tabular}{c}
\includegraphics[angle=0,width=1.0\linewidth]{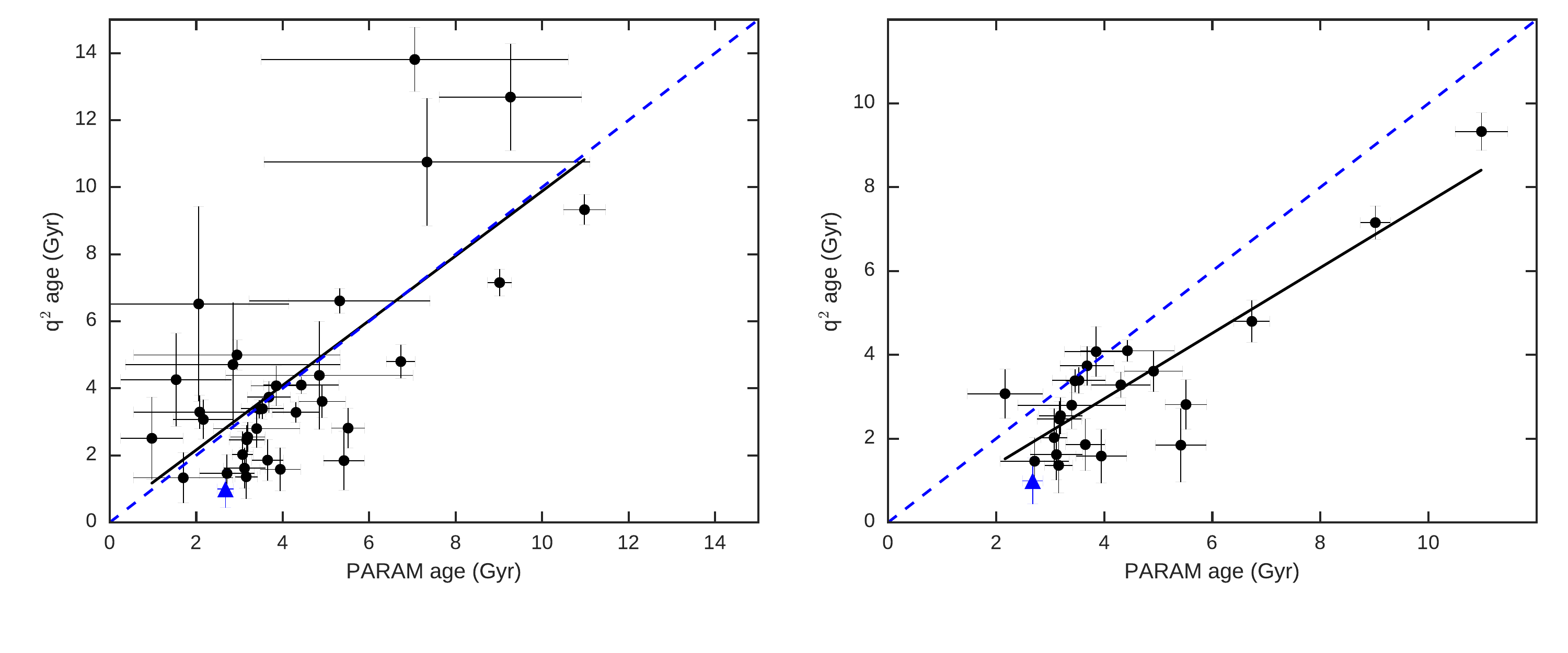}
\end{tabular}
\end{center}
\vspace{-1.1cm}
\caption{Stellar ages derived in this work using the PARAM web interface against the ages derived by the q$^{2}$ package.
The left panel presents all the stars with derived ages, and the right panel shows the stars that have precise age determinations with errors smaller than
1 Gyr in both methods.}
\label{plot_age_age}
\end{figure*}

Elemental abundances for the stars were also determined using an LTE analysis and the same tools and codes as for the stellar parameter determination.
The line list, atomic data, damping parameter, and the detailed description of the abundance derivation is presented in \citet{Adibekyan-16a}.
Again, the EWs of the lines were derived with ARES v2 code with careful visual inspection. In a few cases, when the ARES measurements were not 
satisfactory (this can be caused for instance through cosmics or bad pixels), we measured the EWs using the task \emph{splot} 
in IRAF\footnote{IRAF is distributed by National Optical Astronomy Observatories,
operated by the Association of Universities for Research in Astronomy, Inc., under contract with the National Science Foundation, USA.}. 
The final abundances of the elements (when several spectral lines were available) were calculated as a weighted mean of all
the abundances, where  we used the distance from the median abundance
as a weight \citep{Adibekyan-15}. 
The average of TiI and TiII was used for Ti abundance. The stellar parameters and abundances of the elements are available at the CDS.

The errors of the [X/H] abundances were calculated as a quadratic sum of the errors that are due to EW measurements and those due to uncertainties in the atmospheric
parameters. When three or more lines were available, the EW measurement error was estimated as $\sigma /\sqrt{(n-1)}$, where $\sigma$ is 
the line-to-line abundance-weighted scatter \citep{Adibekyan-15} and $n$ is the number of the observed lines. The errors arising from uncertainties in the stellar parameters 
were calculated as a quadratic sum of the abundance sensitivities on the variation of the stellar parameters by their one-$\sigma$ uncertainties.

\begin{figure*}
\begin{center}
\begin{tabular}{c}
\includegraphics[angle=0,width=0.95\linewidth]{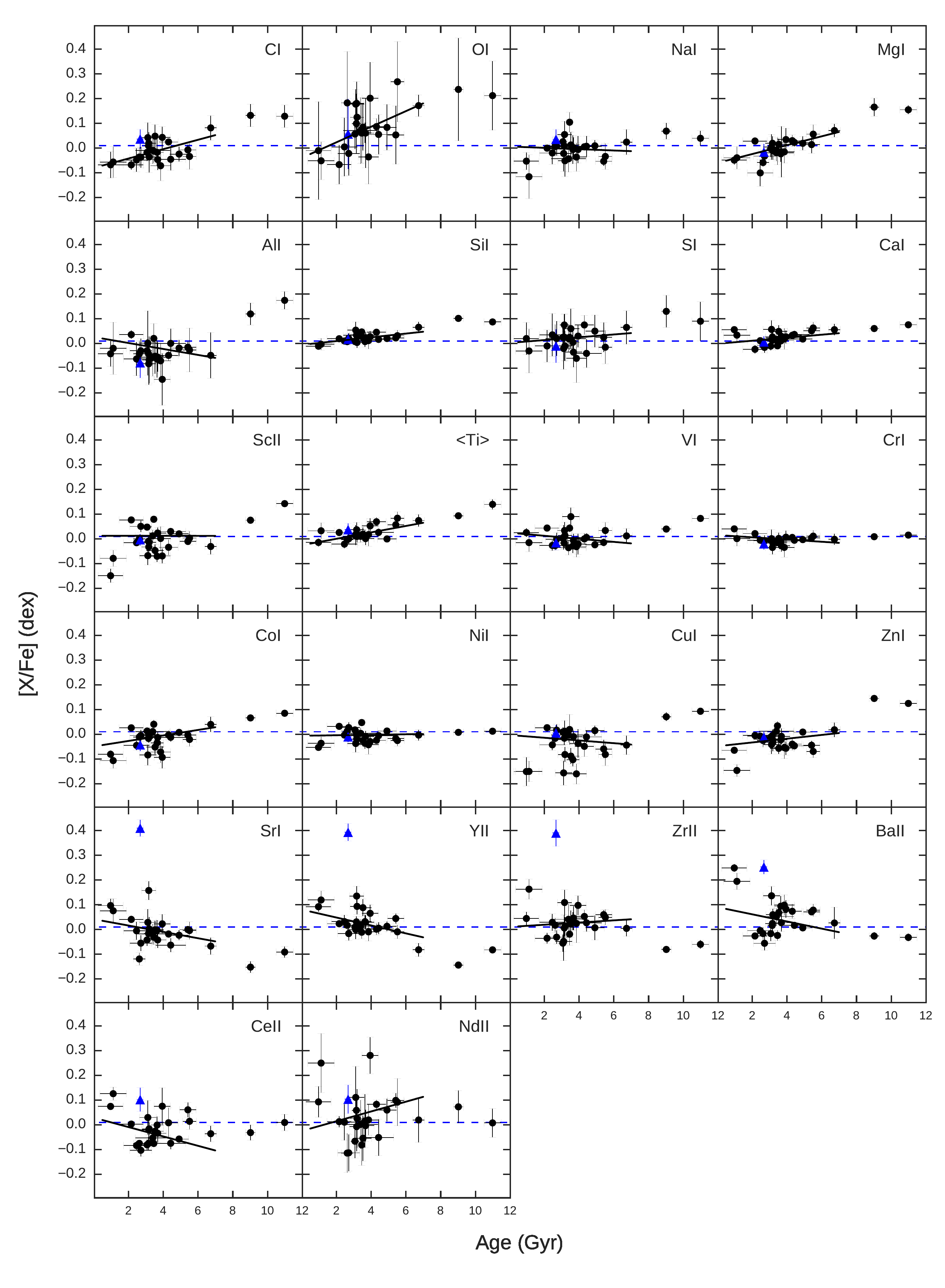}
\end{tabular}
\end{center}
\vspace{-0.9cm}
\caption{Abundance ratio [X/Fe] against age for the stars with the errors in the ages smaller than 1 Gyr. The symbols represent the same as in 
Fig.~\ref{plot_elfe_feh}. The Ba star and the two oldest stars are not considered for the WLS fit.}
\label{plot_elfe_age}
\end{figure*}

\citet{Adibekyan-16a} showed that the $\sigma$ calculated from two lines is usually smaller than the real expected error. These 
underestimated errors can play a crucial role in the incorrect determination of the \tc \ slopes because of their very high weight (if a weighted least-square, WLS,
is used to calculate the best fits). For the abundances of elements that have only two observed spectral lines  (except for oxygen), more realistic errors 
were calculated following the procedure described in \citet{Adibekyan-16a}. In brief, we first calculated the errors in EWs following  \citet[][]{Cayrel-88},
then propagated these to derive the abundance uncertainties for each line. The final uncertainties for the average abundance were 
propagated from the individual errors.

When only one line for a given element was available, as is the case for O (for some stars) and Sr, we determined the error by 
measuring a second EW with the position of the continuum displaced within the root mean square, \textit{rms} 
(due to the noise of the spectra) and by calculating the difference in abundance with respect to the original value. 

The two oxygen lines (6158.2\AA{} and 6300.3\AA{}) used in this work are very weak and deserve special attention \citep[see e.g.][]{Bertrandelis-15}. 
Even when both lines were observed for a spectrum, we calculated individual errors for each line (as described in the previous paragraph) and 
then propagated the error of the average oxygen abundance.

In Fig.~\ref{plot_elfe_feh} we show [X/Fe] abundance trends relative to the stellar metallicity for the sample of stars. The WLS fit of the data is shown 
in the plot by a black line. In the linear regression the inverse of the variance ($\sigma^2$) of the abundances was used as weight. 
Table\,\ref{tab:slopes} presents the slope of the linear dependence and the p-values coming from 
the F-statistics that test the null hypothesis that the data can be modelled accurately by setting the regression coefficients to zero.

In Fig.~\ref{plot_elfe_feh}  a star with strong enhancement in SrI, YII, ZrII, and moderate enhancement in BaII is clearly visible.
The chemical anomalies of this star are reminiscent of  Ba stars \citep[e.g.][]{Bidelman-51,McClure-84}, although these are usually  G-K-giants \citep[e.g.][]{Yang-16},
while our target is warmer and is located in the main sequence. This chemically peculiar star, HD157060, has an astrometric companion with a mass of 
about 0.6M$_\odot$ \citep[][]{Tokovinin-14}. Because of the observed anomalies we did not include this star in our statistical analysis, even though it is shown in the plots.

\begin{figure}
\begin{center}
\begin{tabular}{c}
\includegraphics[angle=0,width=1.0\linewidth]{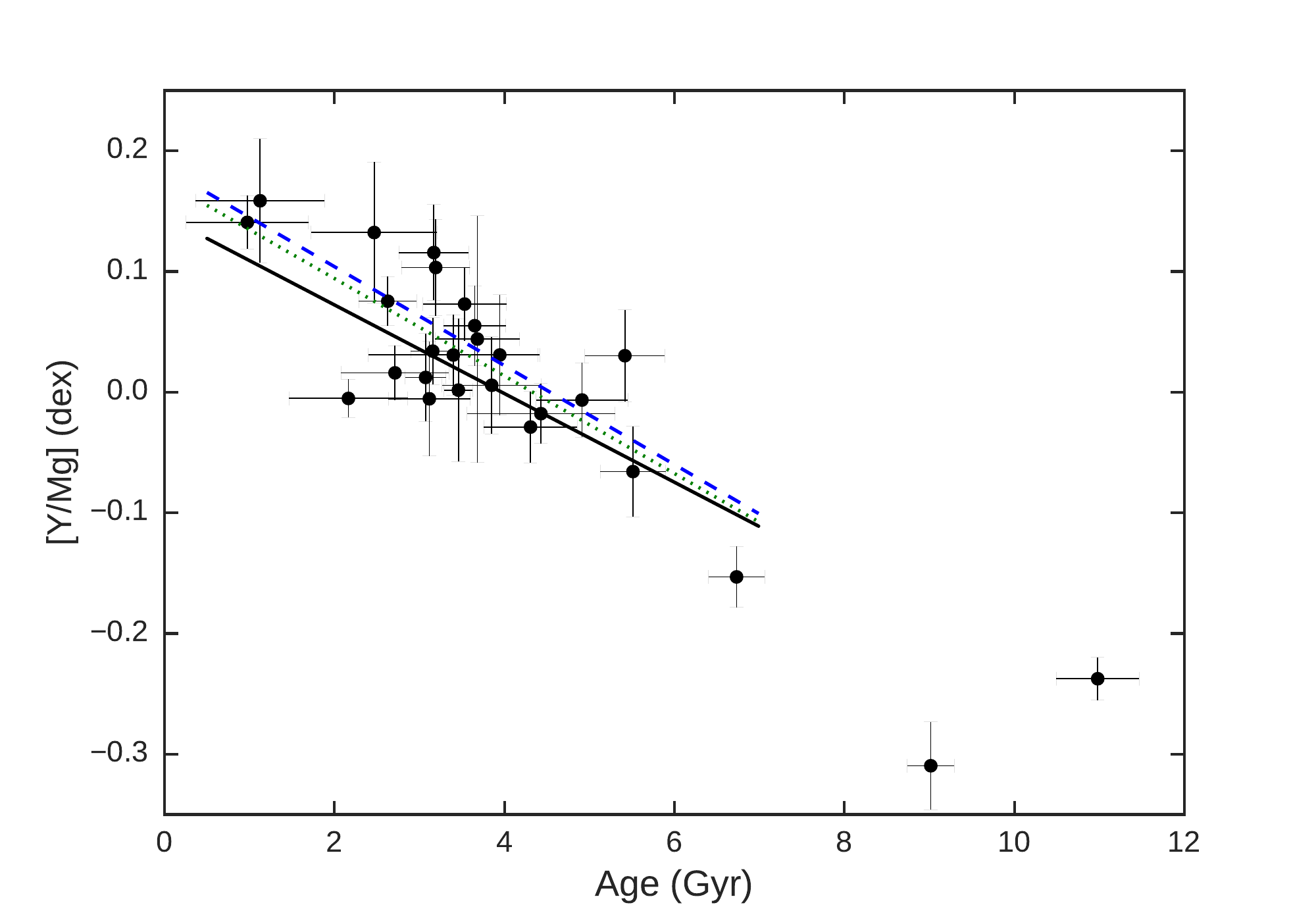}
\end{tabular}
\end{center}
\vspace{-0.5cm}
\caption{[Y/Mg] versus stellar age for stars with age determination errors smaller than one Gyr. The symbols are the same as in the previous plots. 
The black solid line is the linear fit of our data without considering the two oldest stars. 
The blue dashed line and the green dotted lines represent the linear fits from \citet{Nissen-15} and \citet{TucciMaia-16},
respectively.}
\label{plot_y_mg_age}
\end{figure}

\begin{figure}
\begin{center}
\begin{tabular}{c}
\includegraphics[angle=0,width=1.0\linewidth]{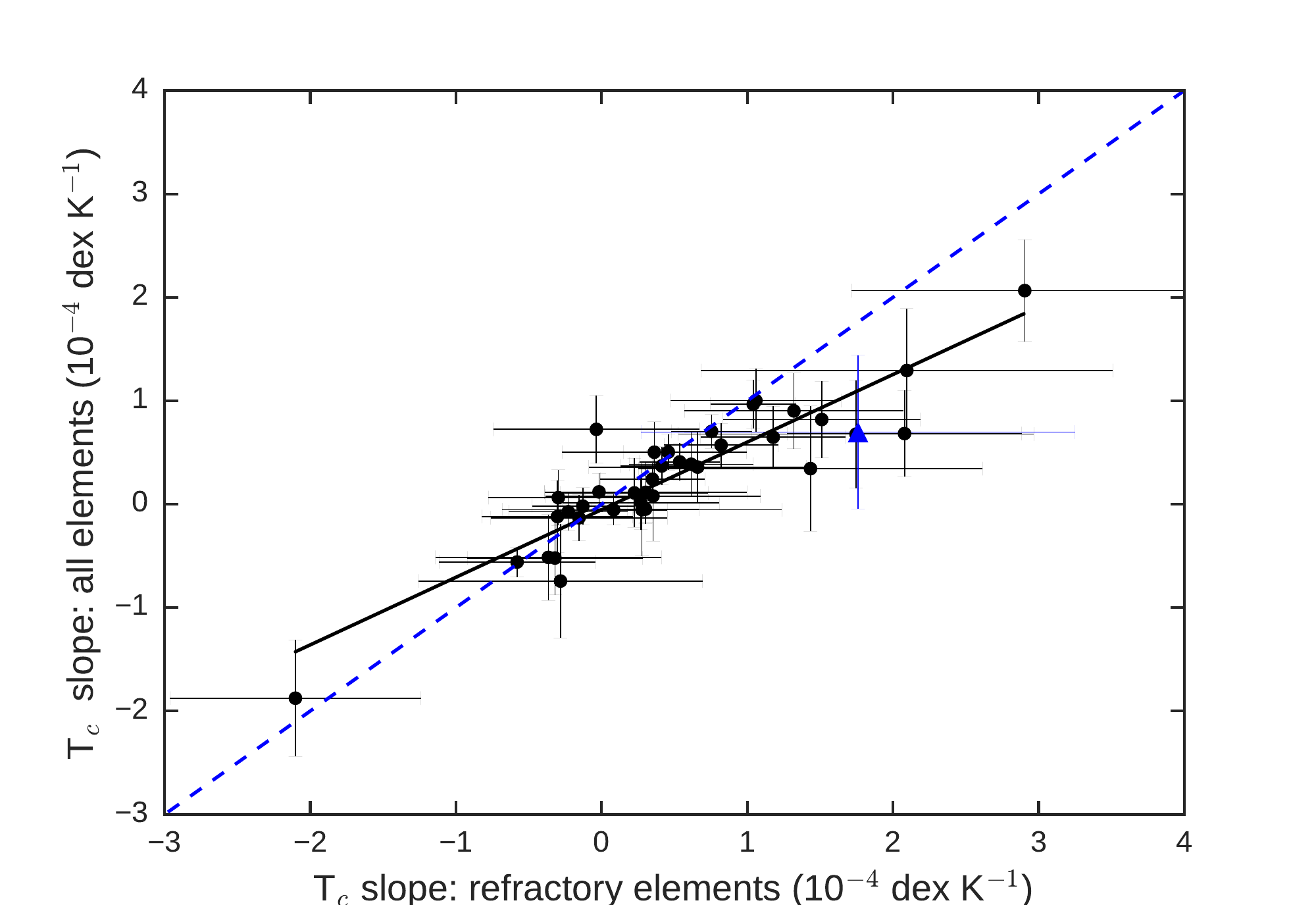}
\end{tabular}
\end{center}
\vspace{-0.5cm}
\caption{\tc \ slopes derived by using all the elements against \tc \ slopes derived by considering only elements that have \tc \ $>$ 900 K.
The black solid line is the linear fit of the data and the blue dashed line represents the one-to-one line.  The symbols are the same as in 
Fig.~\ref{plot_elfe_feh} and the previous plots.}
\label{plot_slope_vs_slope}
\end{figure}

\section{Stellar ages}                                          \label{sec:ages}

In Sect.\,\ref{sec:data} we mentioned that the stars employed in this work were initially selected to have ages very similar to that of our Sun. These ages were derived 
using photometric stellar parameters and BASTI stellar isochrones \citep[][]{Casagrande-11}. We rederived ages of the stars using the new spectroscopic parameters
obtained in this work. We used the PARAM web interface\footnote{http://stev.oapd.inaf.it/cgi-bin/param} with PARSEC isochrones \citep{Bressan-12} 
for a Bayesian estimation of the stellar ages. In addition to \teff \ and [Fe/H], this method also requires the stellar magnitude and the parallax as input parameters. The parallaxes 
of the stars were taken from \citet{vanLeeuwen-07} and the V magnitudes from \citet{Hauck-98}. The ages derived in this way vary from about 1 to 11 Gyr 
(see Table\,\ref{tab:parameters}), however, while as mentioned above they were selected to have ages very similar to that of our Sun. This
once again demonstrates the difficulty of obtaining reliable stellar ages for main-sequence field stars. 

\citet{Melendez-12} claimed that more precise ages can be obtained with the isochrone method when a spectroscopic surface gravity is used as an input
parameter instead of the absolute V
magnitude (derived from the V magnitude and parallax). However, we note that the derivation of 
spectroscopic precise \logg \ is not an easy task. We  used the q$^{2}$ Python package\footnote{The q$^{2}$ source code is available online at
https://github.com/astroChasqui/q2} that uses Yonsei-Yale isochrones \citep{Yi-01,Kim-02} and spectroscopic stellar parameters for the  derivation of stellar ages.
With this method we were able to derive ages for 35 of the 39 stars. For the remaining four stars the code did not provide ages with physical meaning.

\begin{figure*}
\begin{center}
\begin{tabular}{c}
\includegraphics[angle=0,width=1.0\linewidth]{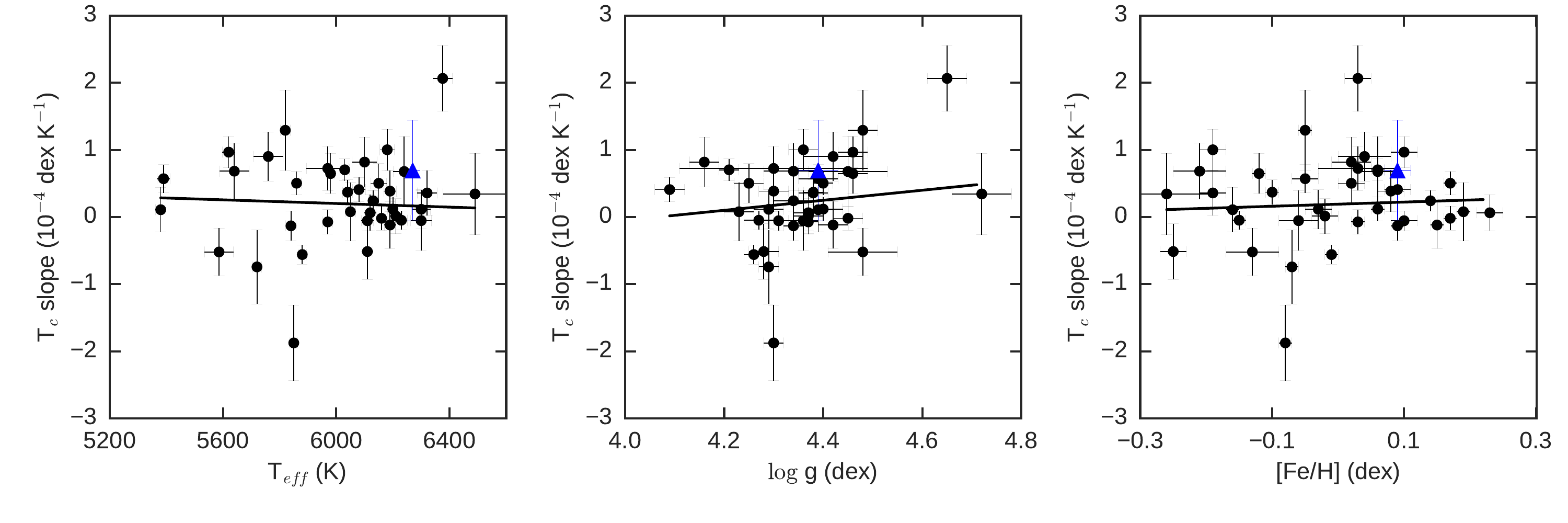}
\end{tabular}
\end{center}
\vspace{-0.5cm}
\caption{\tc \ slopes against stellar parameters. The symbols are the same as in the previous plots.}
\label{plot_slope_param}
\end{figure*}

In the left panel of Fig.~\ref{plot_age_age} we plot stellar ages derived with the PARAM web interface and the q$^{2}$ package. The plot shows that although
there is no offset between the parameters, the scatter is very large. However, the scatter becomes  much smaller when only stars with small errors 
(smaller than 1 Gyr) in both age estimations are considered (right panel of  Fig.~\ref{plot_age_age}). It is very interesting to note that stars with errors on age 
larger than one Gyr mostly lie above the one-to-one line, which
means that ages derived by using spectroscopic \logg \ are usually greater. This produces an offset 
of about one Gyr between the two age estimation methods.  Only stars with precise PARAM ages with errors 
smaller than 1 Gyr are considered in the
remainder of this paper when ages are used. The cut in the error of stellar ages can in principle produce a bias because the parallaxes of the stars that are located farther away
have larger errors. However,  we did not find a strong bias that may change our results significantly for our sample. In particular, we found that the average 
parallaxes are 20$\pm$9 and 25$\pm$19 $mas$ for stars with small and large errors on age, respectively. The ages (3.9$\pm$2.1 and 4.1$\pm$2.5 Gyr) and 
the \rmean \ values (7.7$\pm$1.1 and 7.2$\pm$1.2 kpc) do not deviate dramatically either.

We should note that the comparison shown in Fig.~\ref{plot_age_age} is not a code-to-code comparison, but a comparison
of two methods for deriving stellar ages based on parallax and spectroscopic \logg. The q$^{2}$ package also calculates ages with the use of V magnitude and 
parallaxes. If the same approach were used to derive the stellar ages, then the agreement would be much better, as was demonstrated in \citet{Nissen-15}.

Recently, \citet{Nissen-15} used a sample of 21 solar-twin stars to study the correlation between [X/Fe] elemental abundances with stellar age. For most of the 
elements the authors found strong and significant correlations with ages. The correlations of the [X/Fe] ratios with age were later also confirmed by \citet{Spina-16}.

\begin{figure}
\begin{center}
\begin{tabular}{c}
\includegraphics[angle=0,width=1.0\linewidth]{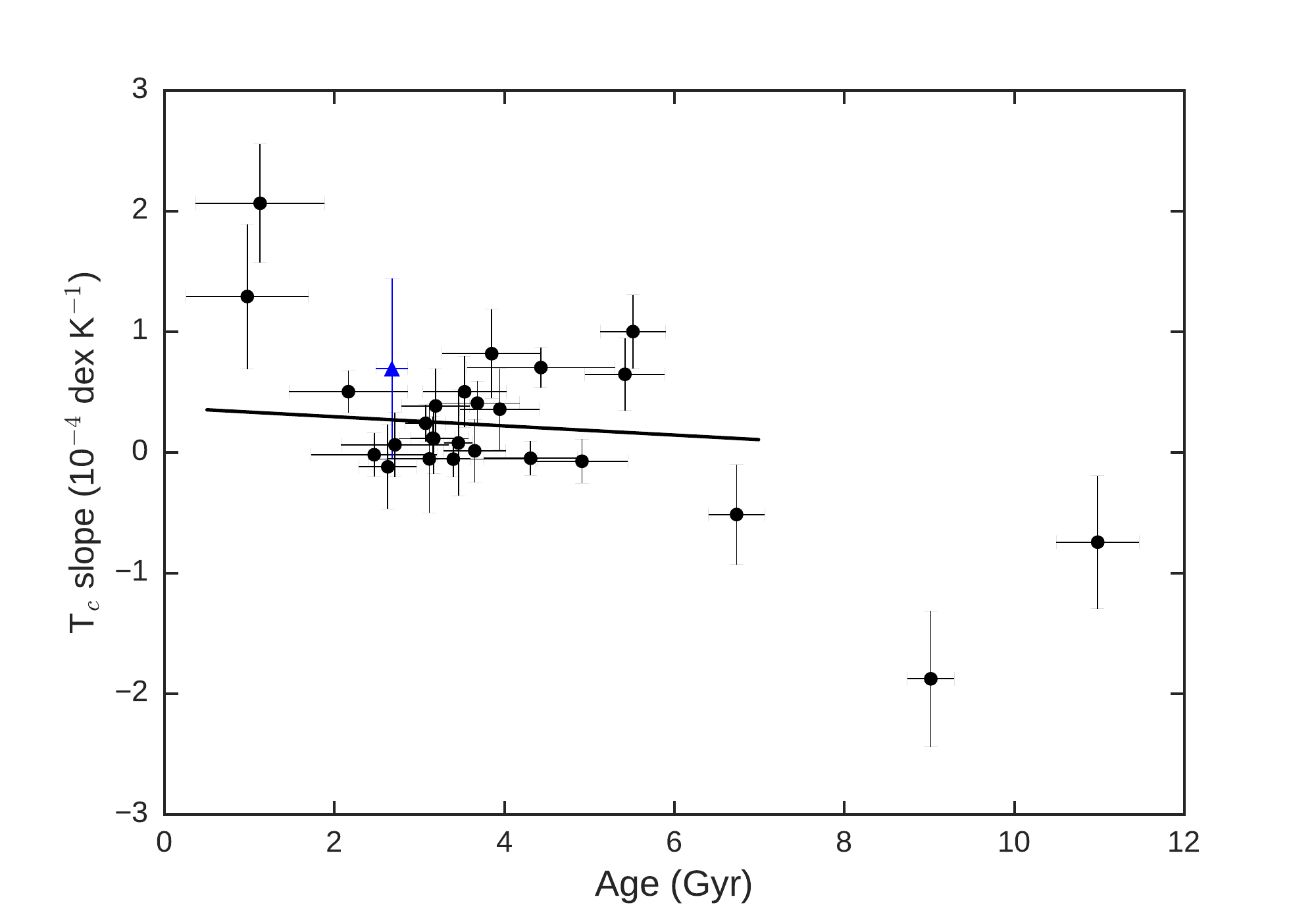}
\end{tabular}
\end{center}
\vspace{-0.5cm}
\caption{\tc \ slopes against stellar age for stars with age determination errors smaller than one Gyr. The two oldest stars and the Ba star are not included in
the fit.} The symbols are the same as in the previous plots.
\label{plot_slope_age}
\end{figure}

In Fig.~\ref{plot_elfe_age} we plot [X/Fe] abundance ratios against stellar ages.  The slopes of the linear WLS fits and the significance of the correlations
are presented in Table\,\ref{tab:slopes}. The two oldest stars are not included in the fit because these older stars might have originated from 
regions with a different spatial homogeneity of metals  that are enriched by supernovae with different neutron excesses (Nissen 2016, submitted).
The table and corresponding plot show that most of the elements show 
significant trends with ages, in a good agreement with \citet{Nissen-15}. The most significant correlations (at a level of about 3$\sigma$) are obtained for
C, O, Mg, Si, Ti, and YII. It is interesting to note that for some elements, 
such as Mg, Si, and Ti (all being $\alpha$-elements), the scatter around the fit is small.

\citet{Nissen-15} showed that [Y/Mg]  can be used to estimate the ages of solar metallicity stars with a precision of about 1 Gyr if a high precision  
of about 0.04 dex in the [Y/Mg] ratio is achieved. This result was later confirmed by \citet{TucciMaia-16}. In Fig.~\ref{plot_y_mg_age} we show the relation 
between [Y/Mg] and stellar ages. The figure confirms the strong and tight correlation between this abundances ratio and age. An alternative way of estimating ages 
of the star would be to apply a multivariate linear regression for the elements that show a strong and significant correlation with age (e.g. Mg, Si, Ti, and Y). 
In this case a better precision in stellar ages is expected because
more elements are used.

\begin{table}[t!]
\setlength{\tabcolsep}{3pt}
\caption{\label{tab:slopes_1} Slopes of the WLS calculus between different pairs of parameters. A frequentist approach is chosen to derive 
the slopes and their uncertainties.}
\begin{center}
\begin{tabular}{lcc}
\hline\hline
 Correlation & Slope$\pm \sigma$ & P(F-stat) \\
\hline
\tc \ slope vs. T$_{eff}$ & -1.35$\pm$3.43 ($\times 10^{-8}$)  & 0.69 \tabularnewline
\tc \ slope vs. $\log g$ & 7.48$\pm$7.81 ($\times 10^{-5}$) & 0.34 \tabularnewline
\tc \ slope vs. {[Fe/H]} & 3.10$\pm$ 6.63 ($\times 10^{-5}$) & 0.64 \tabularnewline
\tc \ slope vs. age & -0.38$\pm$0.79 ($\times 10^{-5}$) & 0.63 \tabularnewline
\tc \ slope vs. R$_{mean}$ & 9.65$\pm$5.93 ($\times 10^{-6}$) & 0.11 \tabularnewline
\tc \ slope vs. R$_{mean}$* & 3.08$\pm$531 ($\times 10^{-6}$) & 0.99 \tabularnewline
\hline
\end{tabular}
\end{center}
Note:$^{(*)}$ Only stars with ages from 2.0 to 4.5 Gyr and with errors below 1 Gyr are considered.
\end{table}

\begin{figure*}
\begin{center}
\begin{tabular}{c}
\includegraphics[angle=0,width=1.0\linewidth]{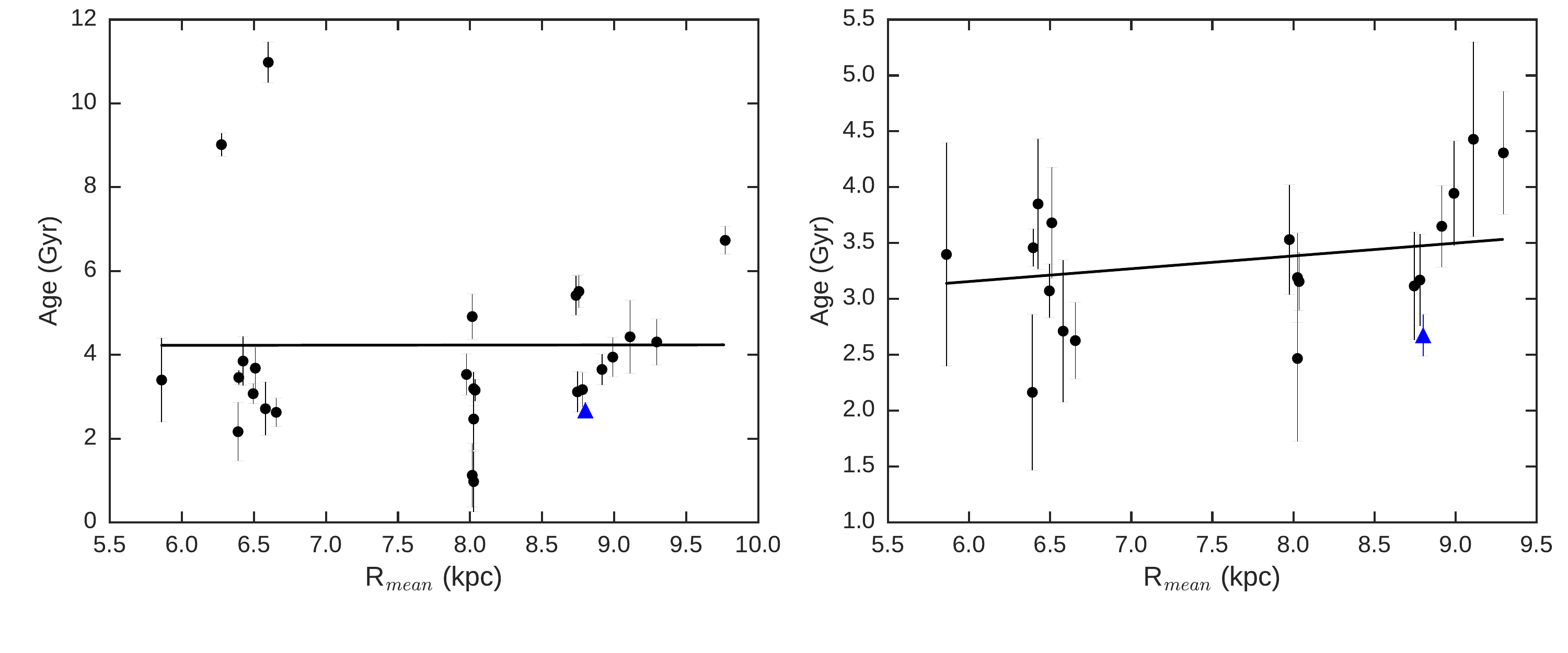}
\end{tabular}
\end{center}
\vspace{-0.5cm}
\caption{Stellar ages again mean Galactocentric distance for all the stars with errors on age smaller than 1 Gyr (left panel). The same as the left panel, but for stars 
with ages from 2.0 to 4.5 Gyr. The symbols are the same as in the previous plots and are described in Fig.~\ref{plot_elfe_feh}.}
\label{plot_age_rmean}
\end{figure*}

\section{\tc \ slope}           \label{sec:tc_slope}

After we derived the differential abundances and corresponding errors, we searched for abundance trends with \tc. 
The 50\%  equilibrium condensation temperatures for a solar system composition gas were taken from \citet{Lodders-03}. 
Although it is common practice to plot [X/Fe]  against \tc \ (and not just [X/H] versus \tc) when searching for a \tc \ trend, this procedure can produce a bias in the derived slope 
\citep{Adibekyan-16a}. The main reason for using [X/Fe] instead of [X/H] is to be able to remove the trends related to 
the Galactic chemical evolution (GCE) from the [X/Fe]--[Fe/H] relations \citep[e.g.][]{GH-13, Saffe-16, Liu-16}. Recently, some authors used the [X/Fe]--age relation
to correct for the GCE \citep[e.g.][]{Yana-16, Spina-16}. However, this correction it is not very simple and straightforward to perform. For example, \citet[][]{Spina-16} introduced
 a correlation between the \tc
\ slope and metallicity after correcting the \tc \ slope for the GCE by using the [X/Fe]--age relation.
The difficulty of correcting for the GCE is probably that the dependence of the [X/Fe] on 
age and metallicity is complex and non-linear. Moreover, the stellar ages and metallicities are also related in a quite complex way as a result of migration
processes in the Galaxy \citep[e.g.][]{Haywood-08a, Haywood-13, Minchev-13, Bergemann-14}. 

We here explore the possible correlation between \tc \ slope and \rmean \, , which, if it exists, should be due to the GCE.
Therefore we did not correct the \tc \ slopes 
for any trends. To avoid any additional biases \citep[see][for details]{Adibekyan-16a}, we also used the linear dependence of [X/H] 
on condensation temperature when deriving the \tc \ trend. In Fig.~\ref{plot_slope_vs_slope} we plot the \tc \ slopes derived when considering all the elements 
against the \tc \ slopes when only refractory elements with \tc \ $>$ 900 K were considered. The plot shows that in general 
the two slope estimates agree, but it also indicates that the slopes derived from refractory elements alone are slightly steeper and the errors are larger. This is 
probably because the condensation temperature range of the refractory elements is much smaller than the full range of the \tc, while the [X/H] range is 
almost the same for refractory and volatile elements. For the remainder of the paper we use the \tc \ slopes derived by considering all the elements.
These slopes are also presented in Table\,\ref{tab:parameters}.

In Figs.~\ref{plot_slope_param} and~\ref{plot_slope_age} we plot the dependence of the \tc \ slope on stellar parameters and on stellar age. The slopes of these
dependencies and their significance levels are presented in Table\,\ref{tab:slopes_1}. The p-values come from the F-statistics that tests the
null hypothesis that the data can be modelled accurately by setting the regression coefficients to zero.
None of the trends with stellar parameters appear significant. At maximum, a $\sim$10\% false-alarm probability was found for the and \rmean versus \tc \ slope correlations.
When the two oldest stars were excluded, our current data do not show any correlation between the \tc \ slope and the stellar age as was observed 
in previous works \citep[e.g.][]{Adibekyan-14, Nissen-15}. This can be due to relatively lower precision in chemical abundances and smaller size of the sample 
compared to the previous works.

\begin{figure*}
\begin{center}
\begin{tabular}{c}
\includegraphics[angle=0,width=0.95\linewidth]{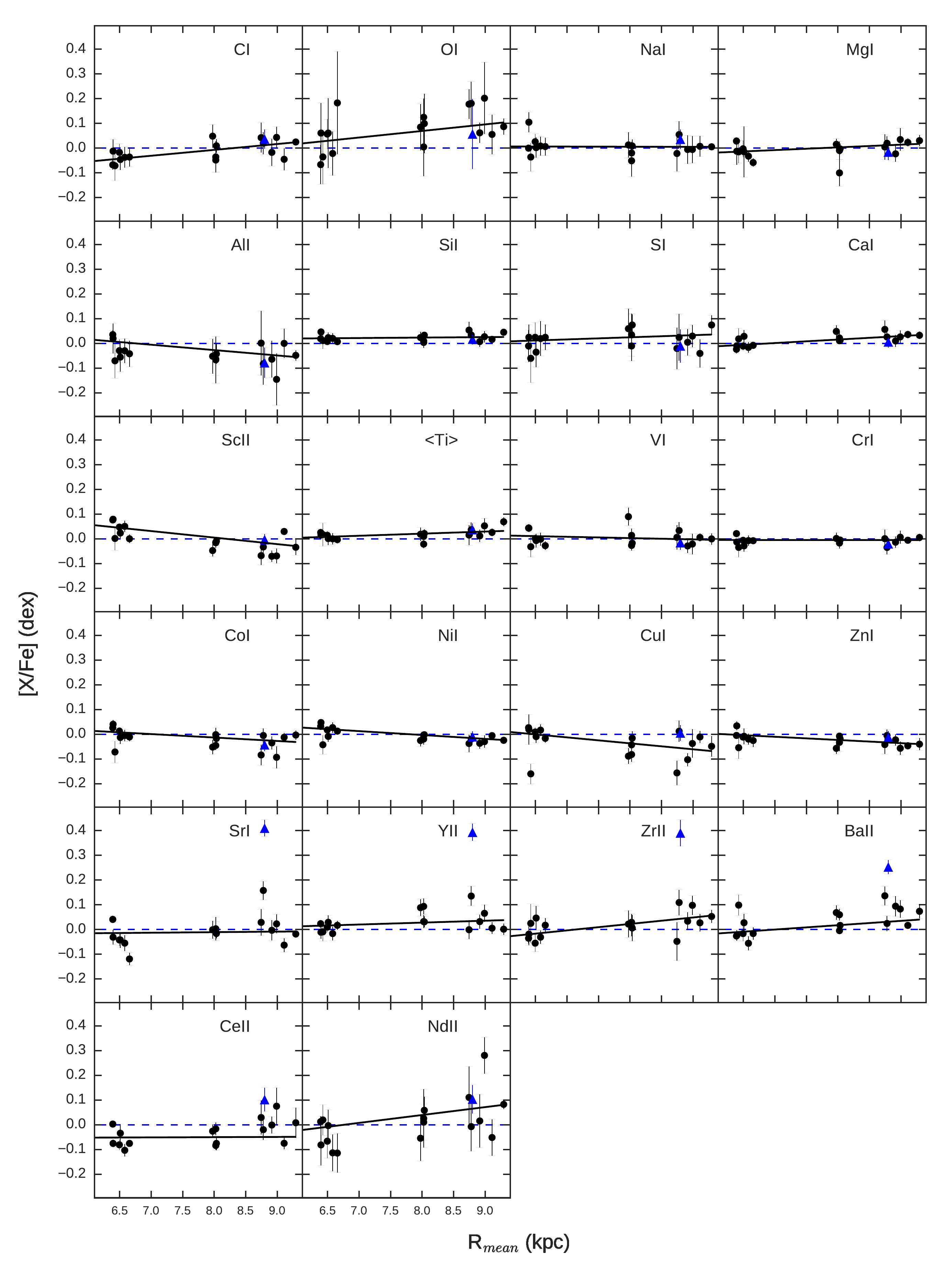}
\end{tabular}
\end{center}
\vspace{-0.5cm}
\caption{Abundance ratio [X/Fe] against \rmean \  for the stars with ages from 2.0 to 4.5 Gyr and with errors on the ages smaller than 1 Gyr. 
The symbols are the same as in Fig.~\ref{plot_elfe_feh}.}
\label{plot_elfe_rmean}
\end{figure*}

\section{GCE and \rmean}                \label{sec:tc_rmean}

Several studies have suggested that the mean of the apo- and pericentric distances (\rmean) is a good indicator of the stellar birthplace 
\citep[e.g.][]{Grenon-87, Edvardsson-93, Nordstrom-99, Rocha-Pinto-04, Haywood-08a, Bensby-13}. However, a word of caution should be added here.  
It has been shown in numerous numerical works, however, that permanent changes in the angular momenta of individual stellar orbits 
(i.e. in their mean or guiding radii) can result from the effect of transient spiral arms \citep{Sellwood-02}, the overlap of bar 
and spiral arms \citep{Minchev-10}, and infalling satellites \citep{Quillen-09}. Because the Milky Way is now well established 
to have a central bar and prominent spiral arms, it is expected that this process, known as radial migration or mixing, has left its footprint 
on the Galactic disk.

\begin{figure*}
\begin{center}
\begin{tabular}{c}
\includegraphics[angle=0,width=0.9\linewidth]{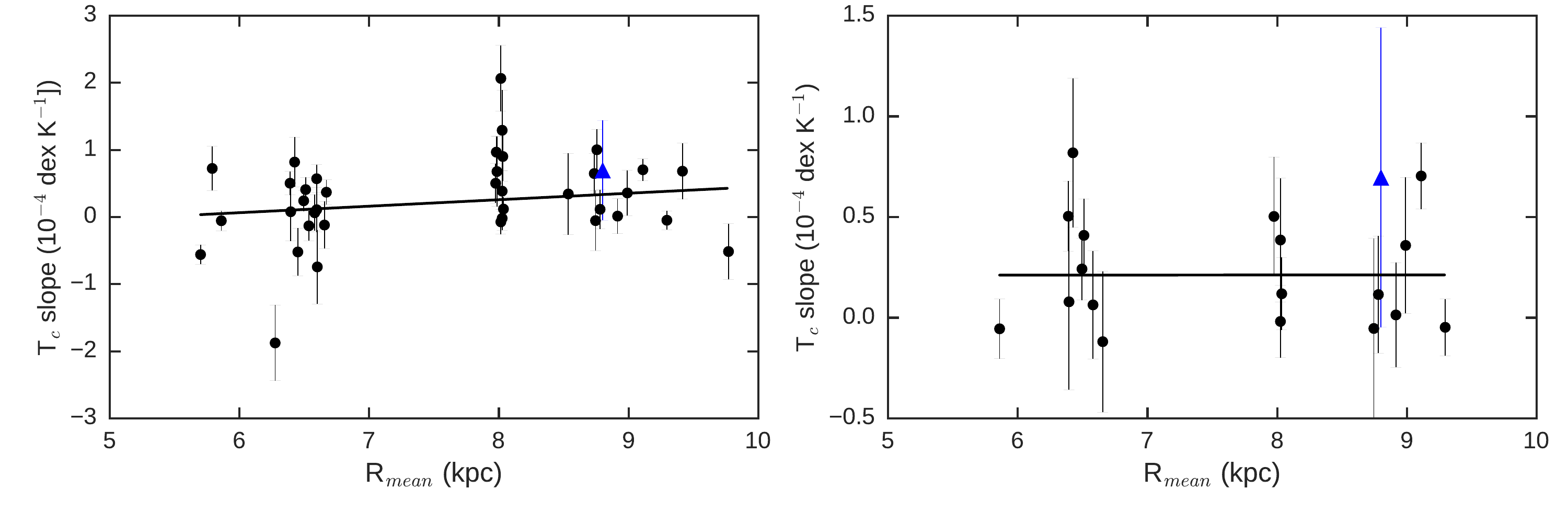}
\end{tabular}
\end{center}
\vspace{-0.5cm}
\caption{\tc \ slope against \rmean \ for all the stars in the sample (left panel) and for stars with precise ages (error smaller than 1 Gyr) from 2.0 to 4.5 Gyr 
(right panel). The symbols are the same as in the previous plots.}
\label{plot_slope_rmean}
\end{figure*}

While the \rmean \ of a star correlates with age \citep[e.g.][]{Rocha-Pinto-04}, the relation  should only be considered in a statistical sense \citep{Wielen-96}.
\cite{Rocha-Pinto-04} showed that young objects (mostly younger than one Gyr)  all have \rmean \ $\approx$ R$_{\odot}$, while older stars present a 
higher proportion of objects coming from different Galactocentric radii. In the left panel of Fig.~\ref{plot_age_rmean} we plot stellar age versus \rmean. 
Although there is no apparent linear correlation between the two parameters, stars with the smallest and largest mean Galactoentric 
distances have ages greater than about two Gyr, and the only two stars with the age of about one Gyr have \rmean \ $\approx$ R$_{\odot}$. 
This result is well in line with those obtained by \citet[e.g.][]{Rocha-Pinto-04}. 
In the right panel of Fig.~\ref{plot_age_rmean} we show only stars with ages from 2.0 to 4.5 Gyr. This is the narrowest age interval where the majority of 
the stars are clustered. The figure shows that the correlation between \rmean \ and stellar age in this age regime is also negligible. 

The \rmean \ values for the sample stars were taken from \citet{Nordstrom-04}. Although the authors did not provide the errors for this parameter, we expect 
an error smaller than 10\%, as was estimated in \citet{Edvardsson-93}.
In Fig.~\ref{plot_elfe_rmean} we plot [X/Fe] abundance ratio as a function of \rmean.
To minimize the effect of stellar age, only stars with ages from 2.0 to 4.5 Gyr
were considered. It is very interesting to see that some elements, such as C, Ca, Ni, and Zr, show statistically significant dependencies on \rmean \
(see Table\,\ref{tab:slopes}). Although our sample is small and the stars show a range of ages (we also recall the difficulties of estimating
precise ages for field main-sequence stars), these results show that different elements show different dependence on \rmean, that is, different radial gradients. 
This result qualitatively agrees with the observations of Galactic abundance gradients by \citet{Lemasle-13}, where the authors used young Galactic 
Cepheids for the gradient derivations, and Galactic abundance gradients obtained from open clusters \citep[e.g.][]{Yong-12, Magrini-15, Cunha-16}.

\subsection{\tc \ slope and \rmean}

It has been shown (see also Sect.\,\ref{sec:parameters}) that abundances of different elements are correlated with the stellar age in different ways. This dependence,
which is due to GCE, produces a correlation between the \tc \ slope and stellar age. It has been also shown that different elements show different dependencies
on the radial distances from the Galactic centre. It is thus logical to expect that this latter dependence, which is also due to Galactic chemical (but not only) 
evolution, may produce a correlation between the \tc \ slope and Galactocentric distances of the stars. 

Following this logic, \citet{Adibekyan-14} found evidence that
the \tc \ slope depends on the mean Galactocentric distances of the stars, which
was  used as a first-order proxy for the birthplace of stars. However, because the stars span a wide range of stellar ages, the authors 
were unable to firmly conclude about the origin of the observed trend. 

In Fig.~\ref{plot_slope_rmean} we show the dependence of the \tc \ slope on \rmean \ of the stars. In the left panel of the plot, where the full sample is plotted,
we can confirm the results of \citet{Adibekyan-14}: stars in the solar circle and in the outer disk have positive slopes, while the stars that formed in the inner
Galaxy show both negative and positive slopes. The mean value (calculated as a weighted arithmetic mean) of the slope for the stars in the inner 
disk (at \rmean \ $\sim$6.5 kpc) is 0.096$\pm$0.438, while in the solar circle (at \rmean \ $\sim$8 kpc) and outer disk (at \rmean \ $\sim$9 kpc) are 0.331$\pm$0.488 and 0.293$\pm$0.396, respectively.
 
In the right panel of Fig.~\ref{plot_slope_rmean} we plot stars in a narrow range of ages (from 2 to 4.5 Gyr) for which the errors on ages are smaller than one Gyr.
The plot and corresponding Table\,\ref{tab:slopes_1} show that the trend completely vanishes. It is difficult to conclude whether the weak correlation
of the \tc \ slope with \rmean \ observed for the full sample is due to a second-order age effect, since the correlation of \rmean \ with age was negligible. We should 
also note that the sample plotted in the right panel of Fig.~\ref{plot_slope_rmean} consists of only 18 stars, which is less than the half of the whole sample, thus 
it suffers (more) from low number statistics.

\section{Summary and conclusion}                \label{sec:conclusion}

To explore the effect of the star formation place (birthplace) on the chemical abundance trend with the condensation temperature,
we selected a sample of main-sequence stars that have similar stellar parameters and expected similar ages as our Sun. The sample consisted of 39 stars,
for 25 of which we carried out new observations with HARPS and UVES spectrographs. The spectra of the remaining 14 stars were taken from HARPS and UVES 
archives (ESO archive). We performed a detailed differential chemical abundance analysis and derived stellar ages for this sample of stars.

When analysing the effect of the GCE on the [X/Fe] elemental ratios, we found that many elements display a strong dependence on stellar metallicity and 
most of the elements show a statistically significant trend with stellar ages. The latter result ([X/Fe] -- age correlation) agrees
well with that of \citet{Nissen-15}.  

To minimize the GCE effect on \tc \ trend that is due to time
or age, we selected a sub-sample of stars that lie in a narrow range of ages from 2 to 4.5 Gyr. The stars were also selected  
to have errors on ages smaller than one Gyr. For this sub-sample of stars we found that the [X/Fe] abundance ratio of some elements correlates with the 
mean of the apo- and pericentric distances, \rmean. These Galactic radial abundance gradients qualitatively agree with those obtained from  
Cepheids \citep[e.g.][]{Lemasle-13} and from open clusters \citep[e.g.][]{Yong-12, Magrini-15}.

When considering only stars that are younger than $\sim$7 Gyr, we found no significant dependence of the \tc \ trend on stellar age, as was observed in previous works \citep[e.g.][]{Adibekyan-14, Nissen-15}. As in \citet{Adibekyan-14}, when using the full sample of stars, we also observed some hint that
the \tc \ slope depends on \rmean, although not in a simple way. However, this dependence completely vanished when a sub-sample of stars
with similar ages (the sub-sample mentioned in the previous paragraph) was considered. Here, we should note that this sub-sample consists of only 
18 stars.

With this small sample we still cannot firmly conclude whether the \tc \ trend depends on the formation place of stars. A larger sample size and 
improvement in the data quality (higher S/N) as well as the use of very similar stars in terms of stellar properties (e.g. metallicity, temperature, age) that 
are observed at a wide range of Galactocentric distances will help to finally answer this question.  In this context, the role of large surveys and missions, such 
as Gaia \citep{Perryman-01}, Gaia-ESO \citep{Gilmore-12}, and APOGEE \citep{Ahn-14}, is invaluable. 

%________________________________________________________________
\begin{acknowledgements}

{This work was supported by Funda\c{c}\~ao para a Ci\^encia e Tecnologia (FCT) through national funds (project ref. UID/FIS/04434/2013) 
and by FEDER through COMPETE2020 (project ref. POCI-01-0145-FEDER-007672). This work was also supported by FCT through the research grants (ref. PTDC/FIS-AST/7073/2014
and ref. PTDC/FIS-AST/1526/2014) through national funds and by FEDER through COMPETE2020 (ref. POCI-01-0145-FEDER-016880 and ref. POCI-01-0145-FEDER-016886).
This work results within the collaboration of the COST Action  TD1308.
P.F., N.C.S., and S.G.S. also acknowledge the support from FCT through Investigador FCT contracts of reference IF/01037/2013, IF/00169/2012, 
and IF/00028/2014, respectively, and POPH/FSE (EC) by FEDER funding through the program ``Programa Operacional de Factores de Competitividade - COMPETE''. 
PF further acknowledges support from (FCT) in the form of an exploratory project of reference IF/01037/2013CP1191/CT0001. 
V.A. and E.D.M acknowledge the support of the FCT (Portugal) in the form of the grants 
SFRH/BPD/70574/2010 and SFRH/BPD/76606/2011, respectively.
V.A also acknowledges the support of COST Action TD1308 through STSM grant with reference Number: COST-STSM-TD1308-32051.
G.I. acknowledges financial support of the Spanish Ministry project MINECO AYA2011-29060.
JIGH acknowledges financial support of the Spanish Ministry of Economy and Competitiveness (MINECO) 
under the 2013 Ram\'{o}n y Cajal program MINECO RYC-2013-14875, and the Spanish ministry project MINECO AYA2014-56359-P.
This research made use of the SIMBAD database operated at CDS, Strasbourg, France.
We thank the referee, Ivan Ram{\'{\i}}rez, for his interesting comments.}

\end{acknowledgements}
%________________________________________________________________

\bibliography{refbib}

\begin{thebibliography}{92}
\expandafter\ifx\csname natexlab\endcsname\relax\def\natexlab#1{#1}\fi

\bibitem[{{Adibekyan} {et~al.}(2016){Adibekyan}, {Delgado-Mena}, {Figueira},
  {Sousa}, {Santos}, {Faria}, {Gonz{\'a}lez Hern{\'a}ndez}, {Israelian},
  {Harutyunyan}, {Su{\'a}rez-Andr{\'e}s}, \& {Hakobyan}}]{Adibekyan-16a}
{Adibekyan}, V., {Delgado-Mena}, E., {Figueira}, P., {et~al.} 2016, \aap, 591,
  A34

\bibitem[{{Adibekyan} {et~al.}(2015{\natexlab{a}}){Adibekyan}, {Figueira}, \&
  {Santos}}]{Adibekyan-16}
{Adibekyan}, V., {Figueira}, P., \& {Santos}, N.~C. 2015{\natexlab{a}},
  [arXiv:1509.02429]

\bibitem[{{Adibekyan} {et~al.}(2015{\natexlab{b}}){Adibekyan}, {Figueira},
  {Santos}, {Sousa}, {Faria}, {Delgado-Mena}, {Oshagh}, {Tsantaki}, {Hakobyan},
  {Gonz{\'a}lez Hern{\'a}ndez}, {Su{\'a}rez-Andr{\'e}s}, \&
  {Israelian}}]{Adibekyan-15}
{Adibekyan}, V., {Figueira}, P., {Santos}, N.~C., {et~al.} 2015{\natexlab{b}},
  \aap, 583, A94

\bibitem[{{Adibekyan} {et~al.}(2012{\natexlab{a}}){Adibekyan}, {Delgado Mena},
  {Sousa}, {Santos}, {Israelian}, {Gonz{\'a}lez Hern{\'a}ndez}, {Mayor}, \&
  {Hakobyan}}]{Adibekyan-12a}
{Adibekyan}, V.~Z., {Delgado Mena}, E., {Sousa}, S.~G., {et~al.}
  2012{\natexlab{a}}, \aap, 547, A36

\bibitem[{{Adibekyan} {et~al.}(2013){Adibekyan}, {Figueira}, {Santos},
  {Mortier}, {Mordasini}, {Delgado Mena}, {Sousa}, {Correia}, {Israelian}, \&
  {Oshagh}}]{Adibekyan-13}
{Adibekyan}, V.~Z., {Figueira}, P., {Santos}, N.~C., {et~al.} 2013, \aap, 560,
  A51

\bibitem[{{Adibekyan} {et~al.}(2014){Adibekyan}, {Gonz{\'a}lez Hern{\'a}ndez},
  {Delgado Mena}, {Sousa}, {Santos}, {Israelian}, {Figueira}, \& {Bertran de
  Lis}}]{Adibekyan-14}
{Adibekyan}, V.~Z., {Gonz{\'a}lez Hern{\'a}ndez}, J.~I., {Delgado Mena}, E.,
  {et~al.} 2014, \aap, 564, L15

\bibitem[{{Adibekyan} {et~al.}(2012{\natexlab{b}}){Adibekyan}, {Santos},
  {Sousa}, {Israelian}, {Delgado Mena}, {Gonz{\'a}lez Hern{\'a}ndez}, {Mayor},
  {Lovis}, \& {Udry}}]{Adibekyan-12b}
{Adibekyan}, V.~Z., {Santos}, N.~C., {Sousa}, S.~G., {et~al.}
  2012{\natexlab{b}}, \aap, 543, A89

\bibitem[{{Ahn} {et~al.}(2014){Ahn}, {Alexandroff}, {Allende Prieto}, {Anders},
  {Anderson}, {Anderton}, {Andrews}, {Aubourg}, {Bailey}, {Bastien}, \&
  et~al.}]{Ahn-14}
{Ahn}, C.~P., {Alexandroff}, R., {Allende Prieto}, C., {et~al.} 2014, \apjs,
  211, 17

\bibitem[{{Beaug{\'e}} \& {Nesvorn{\'y}}(2013)}]{Beauge-13}
{Beaug{\'e}}, C. \& {Nesvorn{\'y}}, D. 2013, \apj, 763, 12

\bibitem[{{Bensby} {et~al.}(2014){Bensby}, {Feltzing}, \& {Oey}}]{Bensby-13}
{Bensby}, T., {Feltzing}, S., \& {Oey}, M.~S. 2014, \aap, 562, A71

\bibitem[{{Bergemann} {et~al.}(2014){Bergemann}, {Ruchti}, {Serenelli},
  {Feltzing}, {Alves-Brito}, {Asplund}, {Bensby}, {Gruyters}, {Heiter},
  {Hourihane}, {Korn}, {Lind}, {Marino}, {Jofre}, {Nordlander}, {Ryde},
  {Worley}, {Gilmore}, {Randich}, {Ferguson}, {Jeffries}, {Micela},
  {Negueruela}, {Prusti}, {Rix}, {Vallenari}, {Alfaro}, {Allende Prieto},
  {Bragaglia}, {Koposov}, {Lanzafame}, {Pancino}, {Recio-Blanco}, {Smiljanic},
  {Walton}, {Costado}, {Franciosini}, {Hill}, {Lardo}, {de Laverny}, {Magrini},
  {Maiorca}, {Masseron}, {Morbidelli}, {Sacco}, {Kordopatis}, \& {Tautvai{\v
  s}ien{\.e}}}]{Bergemann-14}
{Bergemann}, M., {Ruchti}, G.~R., {Serenelli}, A., {et~al.} 2014, \aap, 565,
  A89

\bibitem[{{Bertran de Lis} {et~al.}(2015){Bertran de Lis}, {Delgado Mena},
  {Adibekyan}, {Santos}, \& {Sousa}}]{Bertrandelis-15}
{Bertran de Lis}, S., {Delgado Mena}, E., {Adibekyan}, V.~Z., {Santos}, N.~C.,
  \& {Sousa}, S.~G. 2015, \aap, 576, A89

\bibitem[{{Biazzo} {et~al.}(2015){Biazzo}, {Gratton}, {Desidera}, {Lucatello},
  {Sozzetti}, {Bonomo}, {Damasso}, {Gandolfi}, {Affer}, {Boccato}, {Borsa},
  {Claudi}, {Cosentino}, {Covino}, {Knapic}, {Lanza}, {Maldonado}, {Marzari},
  {Micela}, {Molaro}, {Pagano}, {Pedani}, {Pillitteri}, {Piotto}, {Poretti},
  {Rainer}, {Santos}, {Scandariato}, \& {Zanmar Sanchez}}]{Biazzo-15}
{Biazzo}, K., {Gratton}, R., {Desidera}, S., {et~al.} 2015, \aap, 583, A135

\bibitem[{{Bidelman} \& {Keenan}(1951)}]{Bidelman-51}
{Bidelman}, W.~P. \& {Keenan}, P.~C. 1951, \apj, 114, 473

\bibitem[{{Bond} {et~al.}(2010){Bond}, {O'Brien}, \& {Lauretta}}]{Bond-10}
{Bond}, J.~C., {O'Brien}, D.~P., \& {Lauretta}, D.~S. 2010, \apj, 715, 1050

\bibitem[{{Bressan} {et~al.}(2012){Bressan}, {Marigo}, {Girardi}, {Salasnich},
  {Dal Cero}, {Rubele}, \& {Nanni}}]{Bressan-12}
{Bressan}, A., {Marigo}, P., {Girardi}, L., {et~al.} 2012, \mnras, 427, 127

\bibitem[{{Casagrande} {et~al.}(2011){Casagrande}, {Sch{\"o}nrich}, {Asplund},
  {Cassisi}, {Ram{\'{\i}}rez}, {Mel{\'e}ndez}, {Bensby}, \&
  {Feltzing}}]{Casagrande-11}
{Casagrande}, L., {Sch{\"o}nrich}, R., {Asplund}, M., {et~al.} 2011, \aap, 530,
  A138

\bibitem[{{Cayrel}(1988)}]{Cayrel-88}
{Cayrel}, R. 1988, in IAU Symposium, Vol. 132, The Impact of Very High S/N
  Spectroscopy on Stellar Physics, ed. G.~{Cayrel de Strobel} \& M.~{Spite},
  345

\bibitem[{{Cunha} {et~al.}(2016){Cunha}, {Frinchaboy}, {Souto}, {Thompson},
  {Zasowski}, {Allende Prieto}, {Carrera}, {Chiappini}, {Donor},
  {Garcia-Hernandez}, {Elia Garcia Perez}, {Hayden}, {Holtzman}, {Jackson},
  {Johnson}, {Majewski}, {Meszaros}, {Meyer}, {Nidever}, {O'Connell},
  {Schiavon}, {Schultheis}, {Shetrone}, {Simmons}, {Smith}, \&
  {Zamora}}]{Cunha-16}
{Cunha}, K., {Frinchaboy}, P.~M., {Souto}, D., {et~al.} 2016,
  [arXiv:1601.03099]

\bibitem[{{Dawson} \& {Murray-Clay}(2013)}]{Dawson-13}
{Dawson}, R.~I. \& {Murray-Clay}, R.~A. 2013, \apjl, 767, L24

\bibitem[{{Delgado Mena} {et~al.}(2010){Delgado Mena}, {Israelian},
  {Gonz{\'a}lez Hern{\'a}ndez}, {Bond}, {Santos}, {Udry}, \&
  {Mayor}}]{Delgado-10}
{Delgado Mena}, E., {Israelian}, G., {Gonz{\'a}lez Hern{\'a}ndez}, J.~I.,
  {et~al.} 2010, \apj, 725, 2349

\bibitem[{{Dorn} {et~al.}(2015){Dorn}, {Khan}, {Heng}, {Connolly}, {Alibert},
  {Benz}, \& {Tackley}}]{Dorn-15}
{Dorn}, C., {Khan}, A., {Heng}, K., {et~al.} 2015, \aap, 577, A83

\bibitem[{{Ecuvillon} {et~al.}(2006){Ecuvillon}, {Israelian}, {Santos},
  {Mayor}, \& {Gilli}}]{Ecuvillon-06}
{Ecuvillon}, A., {Israelian}, G., {Santos}, N.~C., {Mayor}, M., \& {Gilli}, G.
  2006, \aap, 449, 809

\bibitem[{{Edvardsson} {et~al.}(1993){Edvardsson}, {Andersen}, {Gustafsson},
  {Lambert}, {Nissen}, \& {Tomkin}}]{Edvardsson-93}
{Edvardsson}, B., {Andersen}, J., {Gustafsson}, B., {et~al.} 1993, \aap, 275,
  101

\bibitem[{{Fischer} \& {Valenti}(2005)}]{Fischer-05}
{Fischer}, D.~A. \& {Valenti}, J. 2005, \apj, 622, 1102

\bibitem[{{Gaidos}(2015)}]{Gaidos-15}
{Gaidos}, E. 2015, \apj, 804, 40

\bibitem[{{Gilmore} {et~al.}(2012){Gilmore}, {Randich}, {Asplund}, {Binney},
  {Bonifacio}, {Drew}, {Feltzing}, {Ferguson}, {Jeffries}, {Micela}, \&
  et~al.}]{Gilmore-12}
{Gilmore}, G., {Randich}, S., {Asplund}, M., {et~al.} 2012, The Messenger, 147,
  25

\bibitem[{{Gonzalez}(1997)}]{Gonzalez-97}
{Gonzalez}, G. 1997, \mnras, 285, 403

\bibitem[{{Gonz{\'a}lez Hern{\'a}ndez} {et~al.}(2013){Gonz{\'a}lez
  Hern{\'a}ndez}, {Delgado-Mena}, {Sousa}, {Israelian}, {Santos}, {Adibekyan},
  \& {Udry}}]{GH-13}
{Gonz{\'a}lez Hern{\'a}ndez}, J.~I., {Delgado-Mena}, E., {Sousa}, S.~G.,
  {et~al.} 2013, \aap, 552, A6

\bibitem[{{Gonz{\'a}lez Hern{\'a}ndez} {et~al.}(2010){Gonz{\'a}lez
  Hern{\'a}ndez}, {Israelian}, {Santos}, {Sousa}, {Delgado-Mena}, {Neves}, \&
  {Udry}}]{GH-10}
{Gonz{\'a}lez Hern{\'a}ndez}, J.~I., {Israelian}, G., {Santos}, N.~C., {et~al.}
  2010, \apj, 720, 1592

\bibitem[{{Grenon}(1987)}]{Grenon-87}
{Grenon}, M. 1987, Journal of Astrophysics and Astronomy, 8, 123

\bibitem[{{Hauck} \& {Mermilliod}(1998)}]{Hauck-98}
{Hauck}, B. \& {Mermilliod}, M. 1998, \aaps, 129, 431

\bibitem[{{Haywood}(2008{\natexlab{a}})}]{Haywood-08}
{Haywood}, M. 2008{\natexlab{a}}, \aap, 482, 673

\bibitem[{{Haywood}(2008{\natexlab{b}})}]{Haywood-08a}
{Haywood}, M. 2008{\natexlab{b}}, \mnras, 388, 1175

\bibitem[{{Haywood} {et~al.}(2013){Haywood}, {Di Matteo}, {Lehnert}, {Katz}, \&
  {G{\'o}mez}}]{Haywood-13}
{Haywood}, M., {Di Matteo}, P., {Lehnert}, M.~D., {Katz}, D., \& {G{\'o}mez},
  A. 2013, \aap, 560, A109

\bibitem[{{Kim} {et~al.}(2002){Kim}, {Demarque}, {Yi}, \& {Alexander}}]{Kim-02}
{Kim}, Y.-C., {Demarque}, P., {Yi}, S.~K., \& {Alexander}, D.~R. 2002, \apjs,
  143, 499

\bibitem[{{Kurucz}(1993)}]{Kurucz-93}
{Kurucz}, R.~L. 1993, {SYNTHE spectrum synthesis programs and line data}

\bibitem[{{Laws} \& {Gonzalez}(2001)}]{Laws-01}
{Laws}, C. \& {Gonzalez}, G. 2001, \apj, 553, 405

\bibitem[{{Lemasle} {et~al.}(2013){Lemasle}, {Fran{\c c}ois}, {Genovali},
  {Kovtyukh}, {Bono}, {Inno}, {Laney}, {Kaper}, {Bergemann}, {Fabrizio},
  {Matsunaga}, {Pedicelli}, {Primas}, \& {Romaniello}}]{Lemasle-13}
{Lemasle}, B., {Fran{\c c}ois}, P., {Genovali}, K., {et~al.} 2013, \aap, 558,
  A31

\bibitem[{{Liu} {et~al.}(2014){Liu}, {Asplund}, {Ram{\'{\i}}rez}, {Yong}, \&
  {Mel{\'e}ndez}}]{Liu-14}
{Liu}, F., {Asplund}, M., {Ram{\'{\i}}rez}, I., {Yong}, D., \& {Mel{\'e}ndez},
  J. 2014, \mnras, 442, L51

\bibitem[{{Liu} {et~al.}(2016){Liu}, {Yong}, {Asplund}, {Ram{\'{\i}}rez},
  {Mel{\'e}ndez}, {Gustafsson}, {Howes}, {Roederer}, {Lambert}, \&
  {Bensby}}]{Liu-16}
{Liu}, F., {Yong}, D., {Asplund}, M., {et~al.} 2016, \mnras, 456, 2636

\bibitem[{{Lodders}(2003)}]{Lodders-03}
{Lodders}, K. 2003, \apj, 591, 1220

\bibitem[{{Mack} {et~al.}(2016){Mack}, {Stassun}, {Schuler}, {Hebb}, \&
  {Pepper}}]{Mack-16}
{Mack}, III, C.~E., {Stassun}, K.~G., {Schuler}, S.~C., {Hebb}, L., \&
  {Pepper}, J.~A. 2016, [arXiv:1601.00018]

\bibitem[{{Magrini} {et~al.}(2015){Magrini}, {Randich}, {Donati}, {Bragaglia},
  {Adibekyan}, {Romano}, {Smiljanic}, {Blanco-Cuaresma}, {Tautvai{\v
  s}ien{\.e}}, {Friel}, {Overbeek}, {Jacobson}, {Cantat-Gaudin}, {Vallenari},
  {Sordo}, {Pancino}, {Geisler}, {San Roman}, {Villanova}, {Casey},
  {Hourihane}, {Worley}, {Francois}, {Gilmore}, {Bensby}, {Flaccomio}, {Korn},
  {Recio-Blanco}, {Carraro}, {Costado}, {Franciosini}, {Heiter}, {Jofr{\'e}},
  {Lardo}, {de Laverny}, {Monaco}, {Morbidelli}, {Sacco}, {Sousa}, \&
  {Zaggia}}]{Magrini-15}
{Magrini}, L., {Randich}, S., {Donati}, P., {et~al.} 2015, \aap, 580, A85

\bibitem[{{Maldonado} {et~al.}(2015){Maldonado}, {Eiroa}, {Villaver},
  {Montesinos}, \& {Mora}}]{Maldonado-15}
{Maldonado}, J., {Eiroa}, C., {Villaver}, E., {Montesinos}, B., \& {Mora}, A.
  2015, \aap, 579, A20

\bibitem[{{Maldonado} \& {Villaver}(2016)}]{Maldonado-16}
{Maldonado}, J. \& {Villaver}, E. 2016, \aap, 588, A98

\bibitem[{{McClure}(1984)}]{McClure-84}
{McClure}, R.~D. 1984, \pasp, 96, 117

\bibitem[{{Mel{\'e}ndez} {et~al.}(2009){Mel{\'e}ndez}, {Asplund}, {Gustafsson},
  \& {Yong}}]{Melendez-09}
{Mel{\'e}ndez}, J., {Asplund}, M., {Gustafsson}, B., \& {Yong}, D. 2009, \apjl,
  704, L66

\bibitem[{{Mel{\'e}ndez} {et~al.}(2012){Mel{\'e}ndez}, {Bergemann}, {Cohen},
  {Endl}, {Karakas}, {Ram{\'{\i}}rez}, {Cochran}, {Yong}, {MacQueen},
  {Kobayashi}, \& {Asplund}}]{Melendez-12}
{Mel{\'e}ndez}, J., {Bergemann}, M., {Cohen}, J.~G., {et~al.} 2012, \aap, 543,
  A29

\bibitem[{{Minchev} {et~al.}(2013){Minchev}, {Chiappini}, \&
  {Martig}}]{Minchev-13}
{Minchev}, I., {Chiappini}, C., \& {Martig}, M. 2013, \aap, 558, A9

\bibitem[{{Minchev} \& {Famaey}(2010)}]{Minchev-10}
{Minchev}, I. \& {Famaey}, B. 2010, \apj, 722, 112

\bibitem[{{Mordasini} {et~al.}(2009){Mordasini}, {Alibert}, {Benz}, \&
  {Naef}}]{Mordasini-09}
{Mordasini}, C., {Alibert}, Y., {Benz}, W., \& {Naef}, D. 2009, \aap, 501, 1161

\bibitem[{{Neuforge-Verheecke} \& {Magain}(1997)}]{Neuforge-97}
{Neuforge-Verheecke}, C. \& {Magain}, P. 1997, \aap, 328, 261

\bibitem[{{Nissen}(2015)}]{Nissen-15}
{Nissen}, P.~E. 2015, \aap, 579, A52

\bibitem[{{Nordstr{\"o}m} {et~al.}(1999){Nordstr{\"o}m}, {Andersen}, {Olsen},
  {Fux}, {Mayor}, {Mowlavi}, \& {Pont}}]{Nordstrom-99}
{Nordstr{\"o}m}, B., {Andersen}, J., {Olsen}, E.~H., {et~al.} 1999, \apss, 265,
  235

\bibitem[{{Nordstr{\"o}m} {et~al.}(2004){Nordstr{\"o}m}, {Mayor}, {Andersen},
  {Holmberg}, {Pont}, {J{\o}rgensen}, {Olsen}, {Udry}, \&
  {Mowlavi}}]{Nordstrom-04}
{Nordstr{\"o}m}, B., {Mayor}, M., {Andersen}, J., {et~al.} 2004, \aap, 418, 989

\bibitem[{{{\"O}nehag} {et~al.}(2014){{\"O}nehag}, {Gustafsson}, \&
  {Korn}}]{Onehag-14}
{{\"O}nehag}, A., {Gustafsson}, B., \& {Korn}, A. 2014, \aap, 562, A102

\bibitem[{{Perryman} {et~al.}(2001){Perryman}, {de Boer}, {Gilmore}, {H{\o}g},
  {Lattanzi}, {Lindegren}, {Luri}, {Mignard}, {Pace}, \& {de
  Zeeuw}}]{Perryman-01}
{Perryman}, M.~A.~C., {de Boer}, K.~S., {Gilmore}, G., {et~al.} 2001, \aap,
  369, 339

\bibitem[{{Quillen} {et~al.}(2009){Quillen}, {Minchev}, {Bland-Hawthorn}, \&
  {Haywood}}]{Quillen-09}
{Quillen}, A.~C., {Minchev}, I., {Bland-Hawthorn}, J., \& {Haywood}, M. 2009,
  \mnras, 397, 1599

\bibitem[{{Ram{\'{\i}}rez} {et~al.}(2015){Ram{\'{\i}}rez}, {Khanal}, {Aleo},
  {Sobotka}, {Liu}, {Casagrande}, {Mel{\'e}ndez}, {Yong}, {Lambert}, \&
  {Asplund}}]{Ramirez-15}
{Ram{\'{\i}}rez}, I., {Khanal}, S., {Aleo}, P., {et~al.} 2015, \apj, 808, 13

\bibitem[{{Ram{\'{\i}}rez} {et~al.}(2009){Ram{\'{\i}}rez}, {Mel{\'e}ndez}, \&
  {Asplund}}]{Ramirez-09}
{Ram{\'{\i}}rez}, I., {Mel{\'e}ndez}, J., \& {Asplund}, M. 2009, \aap, 508, L17

\bibitem[{{Ram{\'{\i}}rez} {et~al.}(2014){Ram{\'{\i}}rez}, {Mel{\'e}ndez}, \&
  {Asplund}}]{Ramirez-14}
{Ram{\'{\i}}rez}, I., {Mel{\'e}ndez}, J., \& {Asplund}, M. 2014, \aap, 561, A7

\bibitem[{{Rocha-Pinto} {et~al.}(2004){Rocha-Pinto}, {Flynn}, {Scalo},
  {H{\"a}nninen}, {Maciel}, \& {Hensler}}]{Rocha-Pinto-04}
{Rocha-Pinto}, H.~J., {Flynn}, C., {Scalo}, J., {et~al.} 2004, \aap, 423, 517

\bibitem[{{Saffe} {et~al.}(2015){Saffe}, {Flores}, \& {Buccino}}]{Saffe-15}
{Saffe}, C., {Flores}, M., \& {Buccino}, A. 2015, \aap, 582, A17

\bibitem[{{Saffe} {et~al.}(2016){Saffe}, {Flores}, {Jaque Arancibia},
  {Buccino}, \& {Jofre}}]{Saffe-16}
{Saffe}, C., {Flores}, M., {Jaque Arancibia}, M., {Buccino}, A., \& {Jofre}, E.
  2016, [arXiv:1602.01320]

\bibitem[{{Santos} {et~al.}(2015){Santos}, {Adibekyan}, {Mordasini}, {Benz},
  {Delgado-Mena}, {Dorn}, {Buchhave}, {Figueira}, {Mortier}, {Pepe},
  {Santerne}, {Sousa}, \& {Udry}}]{Santos-15}
{Santos}, N.~C., {Adibekyan}, V., {Mordasini}, C., {et~al.} 2015, \aap, 580,
  L13

\bibitem[{{Santos} {et~al.}(2001){Santos}, {Israelian}, \& {Mayor}}]{Santos-01}
{Santos}, N.~C., {Israelian}, G., \& {Mayor}, M. 2001, \aap, 373, 1019

\bibitem[{{Santos} {et~al.}(2004){Santos}, {Israelian}, \& {Mayor}}]{Santos-04}
{Santos}, N.~C., {Israelian}, G., \& {Mayor}, M. 2004, \aap, 415, 1153

\bibitem[{{Schuler} {et~al.}(2011{\natexlab{a}}){Schuler}, {Cunha}, {Smith},
  {Ghezzi}, {King}, {Deliyannis}, \& {Boesgaard}}]{Schuler-11a}
{Schuler}, S.~C., {Cunha}, K., {Smith}, V.~V., {et~al.} 2011{\natexlab{a}},
  \apjl, 737, L32

\bibitem[{{Schuler} {et~al.}(2011{\natexlab{b}}){Schuler}, {Flateau}, {Cunha},
  {King}, {Ghezzi}, \& {Smith}}]{Schuler-11}
{Schuler}, S.~C., {Flateau}, D., {Cunha}, K., {et~al.} 2011{\natexlab{b}},
  \apj, 732, 55

\bibitem[{{Sellwood} \& {Binney}(2002)}]{Sellwood-02}
{Sellwood}, J.~A. \& {Binney}, J.~J. 2002, \mnras, 336, 785

\bibitem[{{Smith} {et~al.}(2001){Smith}, {Cunha}, \& {Lazzaro}}]{Smith-01}
{Smith}, V.~V., {Cunha}, K., \& {Lazzaro}, D. 2001, \aj, 121, 3207

\bibitem[{{Sneden}(1973)}]{Sneden-73}
{Sneden}, C.~A. 1973, PhD thesis, THE UNIVERSITY OF TEXAS AT AUSTIN.

\bibitem[{{Sousa}(2014)}]{Sousa-14}
{Sousa}, S.~G. 2014, [arXiv:1407.5817]

\bibitem[{{Sousa} {et~al.}(2015){Sousa}, {Santos}, {Adibekyan}, {Delgado-Mena},
  \& {Israelian}}]{Sousa-15}
{Sousa}, S.~G., {Santos}, N.~C., {Adibekyan}, V., {Delgado-Mena}, E., \&
  {Israelian}, G. 2015, \aap, 577, A67

\bibitem[{{Sousa} {et~al.}(2008){Sousa}, {Santos}, {Mayor}, {Udry},
  {Casagrande}, {Israelian}, {Pepe}, {Queloz}, \& {Monteiro}}]{Sousa-08}
{Sousa}, S.~G., {Santos}, N.~C., {Mayor}, M., {et~al.} 2008, \aap, 487, 373

\bibitem[{{Sozzetti} {et~al.}(2006){Sozzetti}, {Yong}, {Carney}, {Laird},
  {Latham}, \& {Torres}}]{Sozzetti-06}
{Sozzetti}, A., {Yong}, D., {Carney}, B.~W., {et~al.} 2006, \aj, 131, 2274

\bibitem[{{Spina} {et~al.}(2016){Spina}, {Mel{\'e}ndez}, \&
  {Ram{\'{\i}}rez}}]{Spina-16}
{Spina}, L., {Mel{\'e}ndez}, J., \& {Ram{\'{\i}}rez}, I. 2016, \aap, 585, A152

\bibitem[{{Takeda}(2005)}]{Takeda-05}
{Takeda}, Y. 2005, \pasj, 57, 83

\bibitem[{{Takeda} {et~al.}(2001){Takeda}, {Sato}, {Kambe}, {Aoki}, {Honda},
  {Kawanomoto}, {Masuda}, {Izumiura}, {Watanabe}, {Koyano}, {Maehara},
  {Norimoto}, {Okuda}, {Shimizu}, {Uraguchi}, {Yanagisawa}, {Yoshida},
  {Miyama}, \& {Ando}}]{Takeda-01}
{Takeda}, Y., {Sato}, B., {Kambe}, E., {et~al.} 2001, \pasj, 53, 1211

\bibitem[{{Teske} {et~al.}(2015){Teske}, {Ghezzi}, {Cunha}, {Smith}, {Schuler},
  \& {Bergemann}}]{Teske-15}
{Teske}, J.~K., {Ghezzi}, L., {Cunha}, K., {et~al.} 2015, \apjl, 801, L10

\bibitem[{{Teske} {et~al.}(2016){Teske}, {Khanal}, \&
  {Ram{\'{\i}}rez}}]{Teske-16}
{Teske}, J.~K., {Khanal}, S., \& {Ram{\'{\i}}rez}, I. 2016, [arXiv:1601.01731]

\bibitem[{{Thiabaud} {et~al.}(2014){Thiabaud}, {Marboeuf}, {Alibert}, {Cabral},
  {Leya}, \& {Mezger}}]{Thiabaud-14}
{Thiabaud}, A., {Marboeuf}, U., {Alibert}, Y., {et~al.} 2014, \aap, 562, A27

\bibitem[{{Tokovinin}(2014)}]{Tokovinin-14}
{Tokovinin}, A. 2014, \aj, 147, 87

\bibitem[{{Tucci Maia} {et~al.}(2014){Tucci Maia}, {Mel{\'e}ndez}, \&
  {Ram{\'{\i}}rez}}]{TucciMaia-14}
{Tucci Maia}, M., {Mel{\'e}ndez}, J., \& {Ram{\'{\i}}rez}, I. 2014, \apjl, 790,
  L25

\bibitem[{{Tucci Maia} {et~al.}(2016){Tucci Maia}, {Ram{\'{\i}}rez},
  {Mel{\'e}ndez}, {Bedell}, {Bean}, \& {Asplund}}]{TucciMaia-16}
{Tucci Maia}, M., {Ram{\'{\i}}rez}, I., {Mel{\'e}ndez}, J., {et~al.} 2016,
  ArXiv e-prints [\eprint[arXiv]{1604.05733}]

\bibitem[{{van Leeuwen}(2007)}]{vanLeeuwen-07}
{van Leeuwen}, F. 2007, \aap, 474, 653

\bibitem[{{Wielen} {et~al.}(1996){Wielen}, {Fuchs}, \& {Dettbarn}}]{Wielen-96}
{Wielen}, R., {Fuchs}, B., \& {Dettbarn}, C. 1996, \aap, 314, 438

\bibitem[{{Yana Galarza} {et~al.}(2016){Yana Galarza}, {Mel{\'e}ndez},
  {Ram{\'{\i}}rez}, {Yong}, {Karakas}, {Asplund}, \& {Liu}}]{Yana-16}
{Yana Galarza}, J., {Mel{\'e}ndez}, J., {Ram{\'{\i}}rez}, I., {et~al.} 2016,
  \aap, 589, A17

\bibitem[{{Yang} {et~al.}(2016){Yang}, {Liang}, {Spite}, {Chen}, {Zhao},
  {Zhang}, {Liu}, {Liu}, {Liu}, {Deng}, {Spite}, {Hill}, \& {Zhang}}]{Yang-16}
{Yang}, G.-C., {Liang}, Y.-C., {Spite}, M., {et~al.} 2016, Research in
  Astronomy and Astrophysics, 16, 019

\bibitem[{{Yi} {et~al.}(2001){Yi}, {Demarque}, {Kim}, {Lee}, {Ree}, {Lejeune},
  \& {Barnes}}]{Yi-01}
{Yi}, S., {Demarque}, P., {Kim}, Y.-C., {et~al.} 2001, \apjs, 136, 417

\bibitem[{{Yong} {et~al.}(2012){Yong}, {Carney}, \& {Friel}}]{Yong-12}
{Yong}, D., {Carney}, B.~W., \& {Friel}, E.~D. 2012, \aj, 144, 95

\end{thebibliography}

\end{document}